\documentclass{ar-1col}
\RequirePackage[round]{natbib}  

\usepackage{hyperref}
\usepackage{amsmath}  
\usepackage{bm}  

\usepackage{geometry}
\newcommand{\be}{\begin{equation}}
\newcommand{\ee}{\end{equation}}
\newcommand{\ceqn}[1]{equation~\eqref{#1}}

\newcommand{\cfig}[1]{Fig.~\ref{#1}}

\newcommand{\vx}{x}
\newcommand{\vmu}{\mu}

\newcommand{\vDel}{\Delta}

\newcommand{\pparams}{\theta}  
\newcommand{\like}{\mathcal{L}}  
\newcommand{\mlike}{\mathcal{M}}
\newcommand{\namlike}{\mlike^{\rm NA}}  
\newcommand{\drxn}{\bm{\omega}}  
\newcommand{\slike}{\ell}  
\newcommand{\slbl}{\lambda}  
\newcommand{\sset}{S}  
\newcommand{\cset}{\mathcal{C}}  
\newcommand{\Nobj}{N_{\rm obj}}  
\newcommand{\partn}{\varpi}  

\begin{document}

\jname{}
\jvol{}
\jyear{2015}
\doi{}

\title{Probabilistic record linkage in astronomy: Directional cross-identification and beyond}

\markboth%
{Probabilistic record matching in astronomy}
{Budav\'ari \& Loredo}

\author{Tam\'as Budav\'ari$^1$ and Thomas J. Loredo$^2$
\affil{$^1$Department of Applied Mathematics and Statistics, 
The Johns Hopkins University, Baltimore, MD 21218}
\affil{$^2$Center for Radiophysics \& Space Research,
Cornell University, Ithaca, NY 14853}}


\begin{abstract}
Modern astronomy increasingly relies upon systematic surveys, whose dedicated telescopes continuously observe the sky across varied wavelength ranges of the electromagnetic spectrum; some surveys also observe non-electromagnetic ``messengers,'' such as high-energy particles or gravitational waves.
Stars and galaxies look different through the eyes of different instruments, and their independent measurements have to be carefully combined to provide a complete, sound picture of the multicolor and eventful universe.
The association of an object's independent detections is, however, a difficult problem scientifically, computationally, and statistically, raising varied challenges across diverse astronomical applications.
The fundamental problem is finding records in survey databases with directions that match to within the direction uncertainties.
Such astronomical versions of the record linkage problem are known by various terms in astronomy: cross-matching, cross-identification, and directional, positional, or spatio-temporal coincidence assessment.
Astronomers have developed several statistical approaches for such problems, largely independently of related developments in other disciplines.
Here we review emerging approaches that compute (Bayesian) probabilities for the hypotheses of interest:  possible associations, or demographic properties of a cosmic population that depend on identifying associations.
Many cross-identification tasks can be formulated within a hierarchical Bayesian partition model framework, with components that explicitly account for astrophysical effects (e.g., source brightness vs.\ wavelength, source motion, or source extent), selection effects, and measurement error.
We survey recent developments, and highlight important open areas for future research.
\end{abstract}



\maketitle

\section{Introduction}
\label{sec:intro}

The problem of source identification in separate observations is as old as astronomy itself.
When ancient astronomers observed and re-observed a celestial object, they performed a kind of \emph{cross-matching} between previous and new observations by pointing their telescope in a previously recorded direction (after accounting for Earth's rotation) and verifying by eye the identity of the source.
Modern astronomy presents much more challenging cross-matching problems.
The complications are due in part to the huge (and increasing) size of astronomical catalogs, reflecting the increased reach of modern instruments, which have detected hundreds of millions of stars and galaxies densely distributed on the sky.
But the most challenging and most scientifically potent cross-matching problems arise in \emph{multiwavelength astronomy}, where the same patch of sky is observed by instruments probing different parts of the electromagnetic spectrum, and in \emph{multimessenger astronomy}, where one must match electromagnetic observations with observations of other cosmic radiation, such as energetic particles (e.g., cosmic rays or neutrinos) or gravitational waves.

\begin{figure}[t]
\centerline{\includegraphics[width=\textwidth]{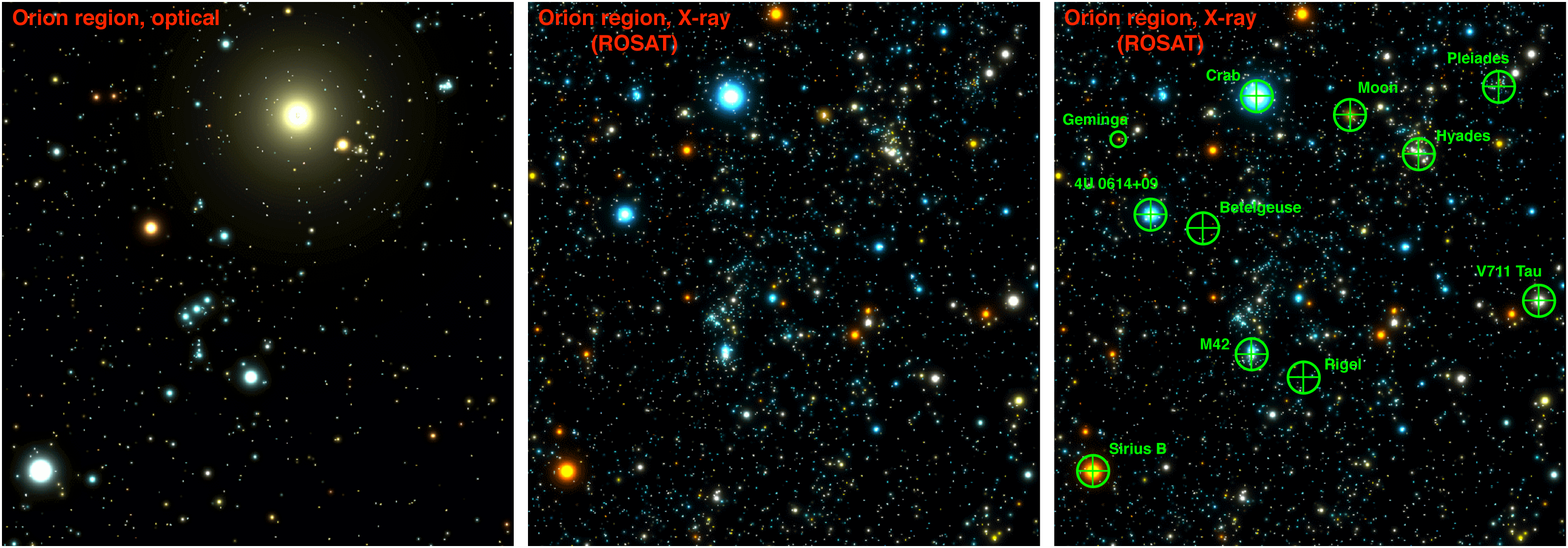}}
\smallskip
\caption{Optical (left) and X~ray (middle) views of the sky in the neighborhood of the Orion constellation; the X~ray image was synthesized using data from \emph{ROSAT} (Dennerl et al.\ 1994).
The right panel repeats the X~ray image, with the locations of notable objects highlighted.}
\label{fig:Orion}
\end{figure}

\cfig{fig:Orion}\ provides a visually accessible example of multiwavelength cross-matching.
The left panel is an optical image of a part of the northern winter night sky that will be familiar even to many non-astronomers:  the region of the constellation Orion.
Besides the familiar stars of Orion's outline, belt, and sword, the field includes a simulated image of the Moon, and other notable objects.
The middle panel shows the same part of the sky, but now as it would appear to X~ray eyes, using data from the \emph{ROSAT} satellite-borne X~ray telescope.%
\footnote{These images were produced by Konrad Dennerl and Wolfgang Voges for a 1994 conference poster; see Dennerl et al.\ (1994) and \url{http://www.mpe-garching.mpg.de/background-picture.html} for details.}
The X~ray image is markedly different from its optical counterpart, and not immediately recognizable as the Orion region.
Notably, some of the brightest and most familiar optically-visible objects, including the giant stars Betelgeuse and Rigel, and the Moon, are dim or invisible in X~rays.
Conversely, many of the brightest X~ray sources are invisible in the optical image.
The right panel repeats the X~ray image, highlighting notable sources, including the Pleides and Hyades open clusters of stars, the Crab pulsar (a spinning neutron star), Sirius~B (a hot white dwarf orbiting the brightest optically visible star, Sirius~A, which is dim in X~rays), Geminga (a pulsar primarily emitting gamma rays), the low-mass X~ray binary 4U~0614+09 (a neutron star accreting matter from a low-mass stellar companion), and the variable star V711~Tau (bright in X~rays because of its unusually hot stellar corona).
The relative brightness (and the colors) of each object conveys a wealth of information about the underlying astrophysics, and can help identify the nature of each object.
But tapping into this information requires reliable cross-matching of sources across the images.

Current and forthcoming surveys are using advanced detectors to image the sky in new regimes and at an unprecedented rate; we expect to have 100 petabytes of image data from astronomical surveys within the next 10 years. 
Direct human examination of such large amounts of data is impossible, hence astronomers implement pipelines to automatically create summary catalogs that contain the measurements of each celestial object.
For example, we record where they are, how bright they appear, as well as their sizes and shapes (reported as estimates with uncertainties).
The surveys produce archives with online interfaces to their catalogs of hundreds of millions of sources. These databases, starting with the Sloan Digital Sky Survey's (SDSS) SkyServer, are profoundly changing the way astronomy is done. 
In the past, most observational astronomers obtained data via direct observation at a telescope facility, often focusing their research on individual objects or small populations.
Today, astronomers increasingly undertake wide-ranging demographic studies of cosmic populations, and much of that starts with accessing remote collections of data in a manner that must link observations from different instruments or epochs.

Observations at different wavelengths require different telescopes and instruments. For example, the night sky is clear from the ground in the visible and radio portions of the electromagnetic spectrum, but only accessible from very high altitudes in other wavebands (e.g., X ray, ultraviolet, infrared or microwave), for which satellites or airborne experiments are needed.
All these measurements have to be combined to infer the physical properties of the observed stars, galaxies and other celestial objects. 
As we study sources far beyond our Milky Way Galaxy, such as distant galaxies, supernovae, or gamma-ray bursts, we can learn about cosmology as the radiation from these objects is modulated by the expansion of the universe.
Distant objects recede from us with speeds increasing with distance, which introduces a reddening in the observed spectrum called redshift.
This is an electromagnetic analog to the Doppler effect for moving sound sources.

When objects can be clearly isolated and precisely localized in all catalogs, and also obviously align to one another, simple procedures can accurately associate observations across catalogs (or at least appear to!).
For example, one can choose to measure pairwise angular separations on the sky, and associate nearest neighbors that are closer than an angular threshold; but ``the devil is in the details.'' 
%
One complication is that the directional measurement error typically differs across catalogs (sometimes dramatically so), and from source-to-source within a catalog, so a simple, fixed threshold is not optimal. 
A further complication is accounting for multiple testing, not just for pairwise matches across two catalogs, but also for matches across three or more catalogs; this can be challenging with conventional approaches relying on Neyman-Pearson style pairwise hypothesis tests.
Finally, the surface density of the sources in separate catalogs can differ by orders of magnitudes.
This coupled with significant directional uncertainties leads to essentially statistical one-to-many and many-to-many linkage problems that are challenging to handle with simple methods.

Finally, surveys in different wavebands typically see different subsets of the objects, and that subset may not be well characterized a priori. 
Typically, each instrument will have a limiting brightness where the sources become indistinguishable from noise. 
The catalogs may list, say, detections with signal-to-noise of five and greater. 
Considering that the spectral energy distribution of each object is different, the brightness limits in different wavebands will typically yield detections of different sets of objects in each survey.

In this paper we review the recent development of probabilistic approaches to tackle the core statistical cross-matching problem, and survey its application across different regimes.
By ``probabilistic'' astronomers typically mean methods that calculate (Bayesian) probabilities for the hypotheses of interest, such as hypothesized associations of particular sources, or demographic properties of a cosmic population that depend on associations (with association uncertainty propagated through the demographic inferences).%
\footnote{For attempts to address some cross-identification problems from a maximum likelihood perspective, see Sutherland and Saunders (1992) and Fioc (2014).}
Section~\ref{sec:bayes} summarizes the key ideas behind the formalism, and Section~\ref{sec:app} reviews its application to a variety of problems in astronomy.
Section~\ref{sec:prac} is concerned with the practical implementation and describes a widely used online service called SkyQuery that can dynamically federate archives of astronomical data.
Section~\ref{sec:sum} concludes by highlighting important open directions for future research.

\section{Hierarchical Bayesian partition models for cross-identification}
\label{sec:bayes}

We distinguish astronomical {\em sources} from astrophysical {\em objects}.
We use ``object'' to refer to a particular astrophysical system.
We use ``source'' to refer to the detection and measurement of an object in an image or other type of astronomical data.
A single object may give rise to multiple sources, with estimated source properties such as direction and brightness reported in one or more {\em catalogs}; alternatively, an object may go undetected for various reasons.
Some objects can produce multiple sources within a \emph{single} image, e.g., the core and lobes of an active galaxy.
The problems we address involve ascertaining whether multiple sources are from the same object; if so, we say the sources are {\em associated}, i.e., we identify the sources with each other (and with a single underlying object).
Such problems are variously called cross-identification, cross-matching, or directional or positional coincidence assessment.\footnote{In astronomical parlance, ``position'' typically refers to the direction toward a source on the celestial sphere.}

A key observable for determining whether sources are associated or not is the source direction (location on the celestial sphere), and for simplicity we focus in this section on cross-identification based solely on directional data.
If directions are measured without uncertainty, and if an object's observable direction does not differ between source measurements (e.g., due to source motion or extent), precise coincidence of source directions would indicate association, and any displacement between sources would rule out association.
But directions are always measured with uncertainty, and this uncertainty makes directional cross-identification nontrivial.
Two sources may have similar directions because of mere coincidence, or because of association; our task is to quantify the evidence for these alternatives.

\subsection{Motivating examples}

Astronomers encounter diverse directional cross-identification tasks.
Before setting out a general probabilistic framework for them, we first quickly survey some examples, to motivate aspects of the framework.
To set the stage, it is helpful to note the following fundamental units and scales arising in astrometry:%
\footnote{\emph{Astrometry} is the measurement of the positions, motions, and apparent brightness of celestial objects.}
\begin{itemize}
\item \emph{Angular units:}
$1\mbox{ rad} \approx 57^\circ$, and $1^\circ = 60\mbox{ arcmin} = 3{,}600\mbox{ arcsec} \approx 0.017\mbox{ rad}$.
The entire celestial sphere has solid angle $4\pi\mbox{ sr} \approx 41{,}000\mbox{ deg}^2 \approx 1.5\times 10^8\mbox{ arcmin}^2$.

\item \emph{Astronomical angular scales:}
The Moon's angular diameter is $\approx 30\mbox{ arcmin} = 0.5^\circ$, corresponding to a
solid angle $\approx 0.2\mbox{ deg}^2$, which is $\approx 4.8\times 10^{-6}$ of the full sky (so it takes $\approx 208,000$ lunar disks to cover the sky).
For ground-based observing, blurring and twinkling due to atmospheric effects  limit the resolution of point sources to about $\sim 1\mbox{ arcsec}$.
The \emph{Hubble Space Telescope} (\emph{HST}), unhindered by the atmosphere, has an angular resolution of $\approx 0.1\mbox{ arcsec}$.
Bright sources can be localized with greater precision than these resolution limits suggest, depending on how accurately the instrumental and atmospheric blurring can be characterized.

\item \emph{Sky densities:}
There are $\approx 6000$ stars visible to the naked eye in dark skies (i.e., brighter than about 6th magnitude in the visible band), corresponding to an average density of $\approx 0.15\mbox{ deg}^{-2}$, or about $0.03/\mbox{lunar disk}$.
Of course, there are many more fainter stars; for example, the density of stars down to 15th magnitude (i.e., with about $1/4000$ the flux of 6th magnitude stars) is about $10^3\mbox{ deg}^{-2}$, and \emph{HST} can detect stars to about 25th magnitude, for which the density is about $60{,}000\mbox{ deg}^{-2}$.
The density of galaxies in the Hubble Ultra-Deep Field is $\approx 0.25 \mbox{ arcsec}^{-2}$ or about $3\times 10^6\mbox{ deg}^{-2}$, corresponding to $\sim 10^{11}$ galaxies over the whole sky.
These are average densities; galaxies are distributed roughly isotropically over the sky, but the stellar density is much greater in the plane of the Milky Way than at the Galactic poles.
\end{itemize}

\begin{figure}[t]
  \centering
  \includegraphics[width=\textwidth]{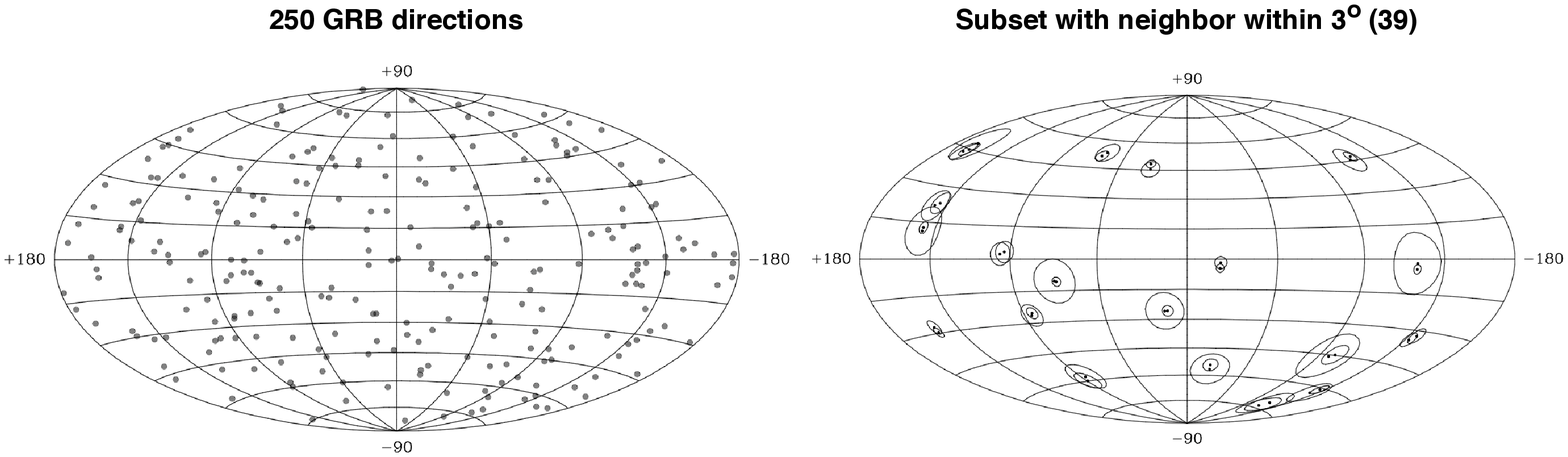}
  \caption{Directions to GRBs detected by BATSE, plotted on an equal-area projection of the celestial sphere using Galactic coordinates (so the equatorial plane is the Galactic plane).
  {\em Left}: Best-fit directions for the first 250 BATSE GRBs.
  {\em Right}:  Best-fit directions and 68\% confidence regions for the 39 GRBs among the 250 that have best-fit directions within 3$^\circ$ of the nearest neighbor.}
  \label{fig:GRB250}
\end{figure}

\cfig{fig:GRB250}\ (left panel) shows the best-fit directions to the first 250 of the $\approx 2700$ gamma-ray bursts (GRBs) detected by the Burst and Transient Source Experiment (BATSE) on the {\em Compton Gamma Ray Observatory}.
GRBs are cosmic explosions discovered by satellite-born gamma ray detectors in the 1960s.
To constrain theories for the objects producing the bursts, astronomers were interested in whether bursts were unique, associated with catastrophic events, or if instead the host system could produce multiple GRBs.
The right panel shows the direction estimates for 39 GRBs from the set of 250 that have nearest neighbors within 3$^\circ$, a smaller angular scale than the uncertainties, which range from 5$^\circ$--25$^\circ$.
Is this a surprising number, providing evidence for GRB repetition (and thus ruling out a catastrophic origin)?
An assessment must account for both the direction uncertainties and the number of GRBs, since as the catalog grows in size, more ``mere coincidence'' candidate matches will appear by chance.
For example, among the first 1000 GRBs, there are 485 such coincident pairs; among the complete catalog of $\approx 2700$ GRBs, there are $\approx 2280$ (i.e., 84\%) such coincident pairs.
Notably, in this regime uncertainty quantification that goes beyond threshold-based matching is crucial; one could have significant evidence for associations without having compelling evidence for any particular association.

The GRB repetition problem, requiring assessment of directional coincidences within a {\em single} catalog, is the simplest kind of directional cross-identification problem arising in astronomy.
It is an astronomical analog to the ``de-duplication'' problem in database record linkage.
It is the first problem in astronomy for which probabilistic cross-identification methods were developed of the type we review here.
Luo, Loredo, and Wasserman (1996; LLW96) and Graziani and Lamb (1996) independently considered spatial and spatio-temporal coincidence models and showed that the BATSE data for the $\approx 1100$ GRBs detected at that time did not favor repeating models (i.e., the posterior probability for the number of genuine repeats was largest for zero repeats).
Other evidence later indicated that GRBs are in fact produced by the catastrophic collapse of massive stars into black holes, producing the most energetic explosions known to astronomers, visible from distant galaxies.
Such events by their very nature cannot repeat (although delayed reobservation via gravitational lensing is a remote possibility).

\begin{figure}[t]
\centerline{\includegraphics[width=.95\textwidth]{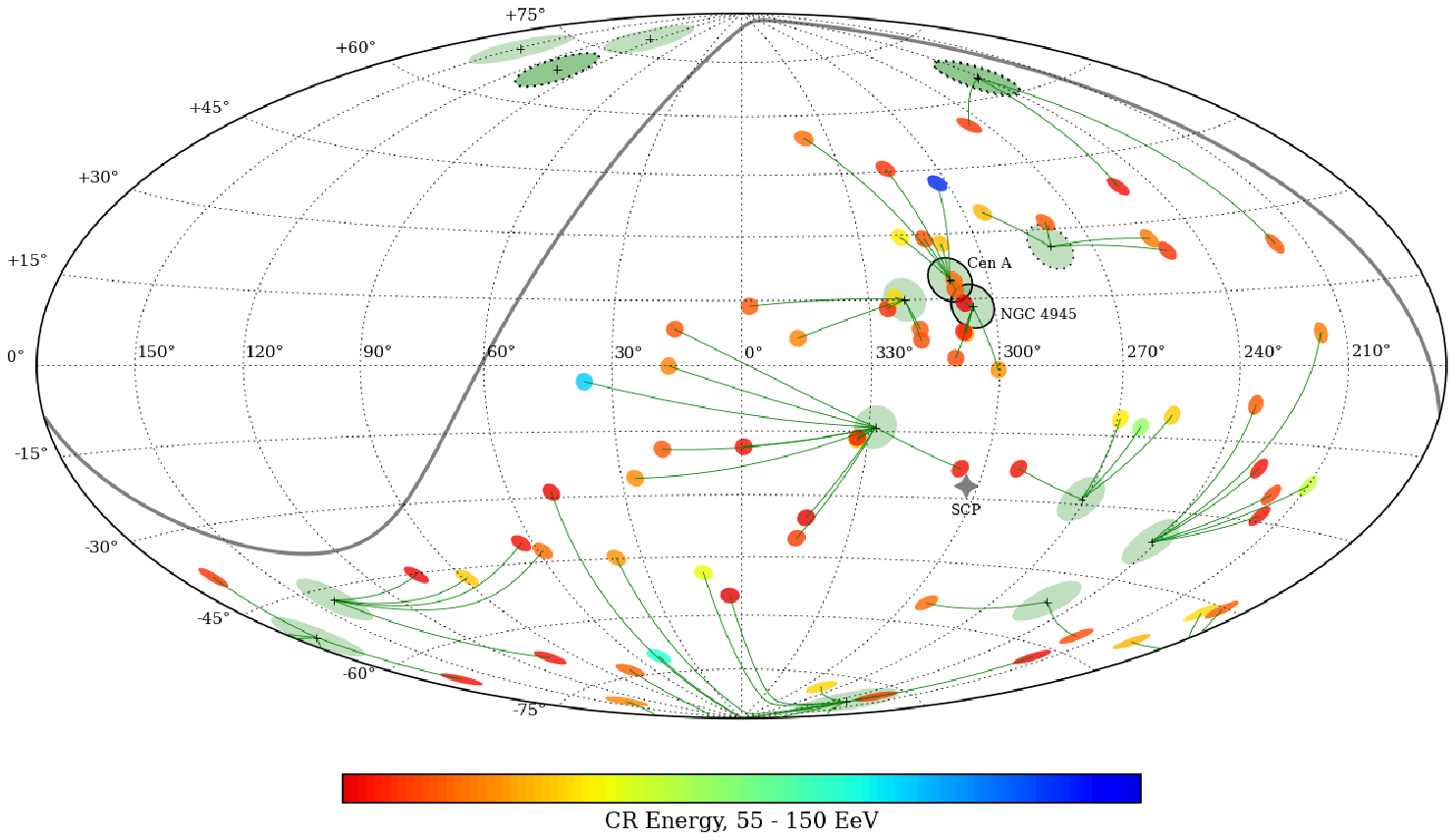}}
\caption{Sky map showing directions to 69 UHECRs detected by PAO, and
to 17 nearby AGN.  Directions are
shown in an equal-area projection in Galactic coordinates.
Thick gray line indicates the boundary of the PAO field of view.
Small tissots show UHECR directions; tissot radius is $2^\circ$ corresponding
to $\approx 2$ standard deviation errors; tissot color indicates UHECR energy.
Large green tissots indicate AGN directions; tissot radius is $5^\circ$
indicating a plausible cosmic ray magnetic deflection scale.
Thin curves are geodesics connecting each UHECR to its nearest AGN.}
\label{fig:UHECR}
\end{figure}

GRBs also motivated the first attempts at between-catalog probabilistic cross-identification, in the search for galaxies that host GRBs (Band and Hartmann 1998).
This was an example of {\em asymmetric} cross-identification:  a catalog of galaxies contains candidate {\em hosts} for the members of the GRB {\em target} population, and one seeks to estimate the fraction of targets that have hosts in the candidate host population, or the fraction of the candidate host population that actually do host a target.
A more recent example along these lines concerns identifying the sites producing ultra-high energy cosmic rays (UHECRs), energetic particles of cosmic origin, with individual particle energies a million or more times greater than can be reached with particle accelerators such as the Large Hadron Collider.
\cfig{fig:UHECR} shows best-fit directions and uncertainties for 69 UHECRs detected by the Pierre Auger Observatory (PAO), along with directions to 17  active galactic nuclei (AGN) near our Galaxy.  AGN are the cores of galaxies harboring supermassive black holes; they are leading candidate sources for UHECRs.
Each UHECR has a small tissot\footnote{A tissot is a map projection of a circle on the sphere, originally introduced as a tool for visualizing map projection distortion; see \url{http://wordpress.mrreid.org/2011/06/07/tissots-indicatrix/}.} indicating uncertainty in its measured direction of arrival.
Each AGN direction has a larger tissot indicating uncertainty in the predicted arrival direction for UHECRs originating from that AGN; the uncertainty is due, not to measurement error, but rather to stochastic deflection of the cosmic ray due to cosmic magnetic fields.

Soiaporn et al.\ (2013) developed a hierarchical Bayesian framework for studying candidate host-UHECR associations, including levels describing cosmic ray emission (e.g., spectrum and event rate), propagation (including deflection), and detection and measurement (with measurement error).
For the nearby AGN host population they studied, evidence for association was equivocal (posterior odds for most models vs.\ models considering UHECRs came from an unspecified, isotropically distributed host population were near unity).
Presuming association, the analysis enables estimation of the UHECR production rate and magnetic deflection scales.

The combination of large and heteroskedastic direction uncertainties for GRBs and UHECRs, and the modest size of the catalogs, motivated the probabilistic analyses just described.
Large uncertainties mean that chance coincidences are not too rare; sound coincidence assessment thus requires careful accounting and propagation of direction uncertainty, which is straightforward within the probabilistic approach (via marginalization over latent object direction).
Implementation formally requires sums over all possible partitions of the data into candidate associations; for catalogs of modest size, it is fairly straightforward to explore the space of plausible candidate associations.

\begin{figure}[t]
\centerline{\includegraphics[width=.7\textwidth]{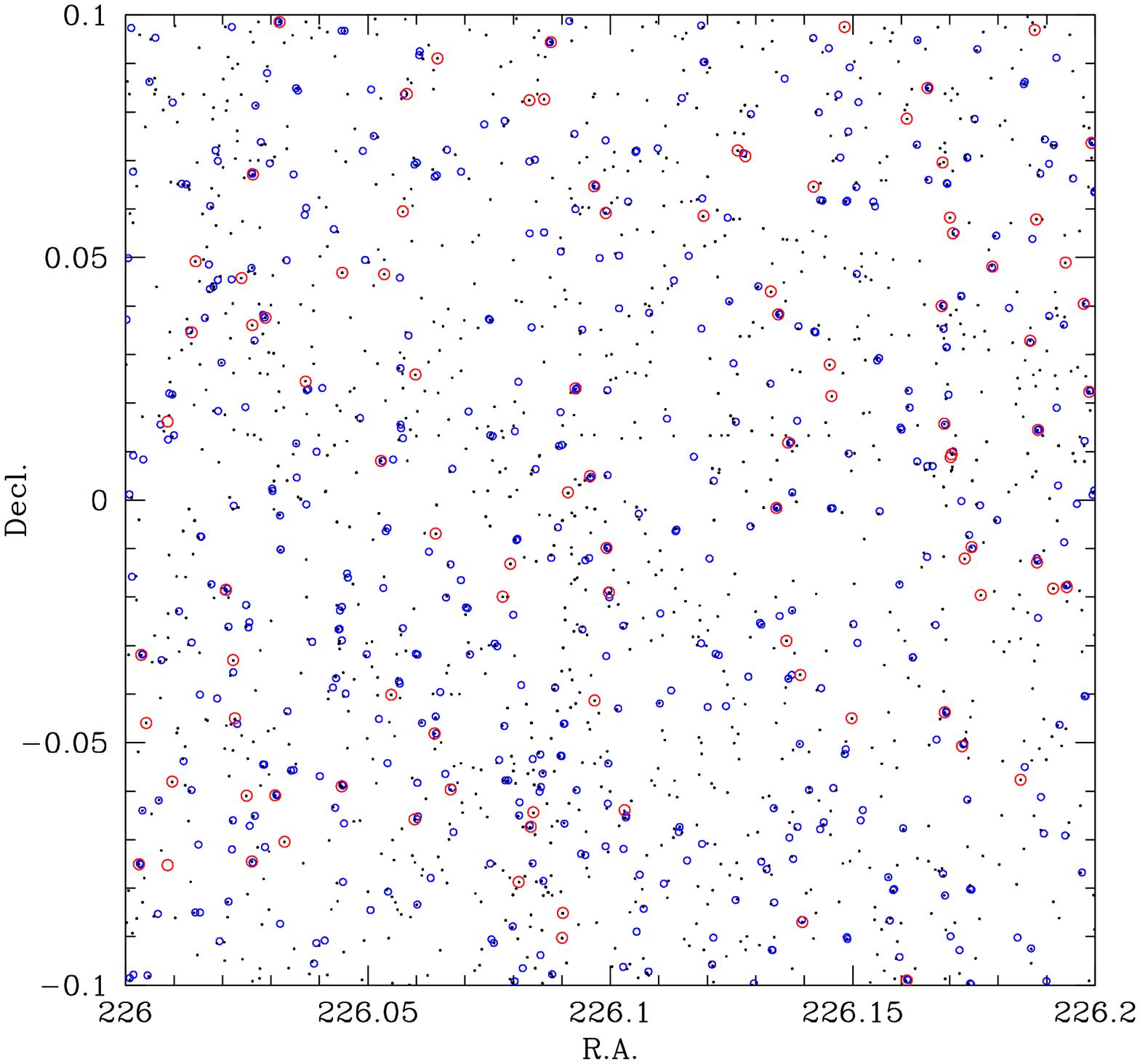}}
\caption{Observations in different wavebands detect different subsets of the astronomical objects in the overlapping field of view.
Shown are best-fit directions for sources in three surveys in a small area of the sky, using equatorial celestial coordinates (right ascension and declination).
Only a fraction of the optical SDSS sources (black dots) have ultraviolet counterparts (blue circles) in the GALEX survey, and there are often ambiguities, where multiple potential counterparts have similar separations.
Infrared sources from the shallower 2MASS catalog (red circles) often have SDSS counterparts, but not always.
Probabilistic cross-identification must account for the different ways each survey samples the underlying population.}
\label{fig:B11-XMatch}
\end{figure}

In contrast to the within-catalog and asymmetric coincidence assessment tasks in the GRB and UHECR problems, the most common cross-identification problems arising in astronomy require \emph{symmetric} matching of sources across two or more large catalogs of sources with small direction uncertainties.
Despite small uncertainties, the large sizes of many star and galaxy catalogs results in significant probabilities for chance coincidences, motivating a probabilistic approach to carefully quantify the evidence for candidate associations.
Such problems are arising with increasing frequency as {\em multiwavelength astronomy} becomes a dominant mode of study, requiring astronomers to match sources between, say, optical, infrared, radio, and X~ray catalogs.\footnote{The emerging field of {\em multimessenger astronomy}---combining electromagnetic observations with observations of particles or gravitational waves---will provide further such problems.}
Budav\'ari and Szalay (2008; BS08) pioneered development of probabilistic methods for symmetric cross-identification in this regime.
\cfig{fig:B11-XMatch} shows an example of symmetric cross-identification of sources in optical (Sloan Digital Sky Survey: SDSS), infrared (Two Micron All Sky Survey: 2MASS), and ultraviolet (GALEX) catalogs (Budav\'ari 2012).
The catalogs contain many millions of sources; the team implemented large-scale Bayesian cross-identification using graphics processing units (GPUs) to make the  approach feasible at this scale.
These algorithms have made large-scale symmetric cross-matching routine and transparent; many astronomers rely on these algorithms unaware of the statistical and computational techniques underlying their database queries.

\subsection{Statistical elements}

The probabilistic approach to cross-identification is built using a hierarchical Bayesian formalism.
Tancredi and Liseo (2011), Steorts,  Hall, and Fienberg (2014), and Sadinle (2014) have recently developed similar approaches for database record linkage problems, and the motivations they cite carry over to the astronomical setting:  the ability to quantify uncertainty in matches; the ability to straightforwardly yet thoroughly quantify and propagate uncertainties (e.g., due to measurement error or attribute reliability); the flexibility of graphical (hierarchical) modeling for accounting for rich generative model structure (as in the UHECR example); and the ability to easily handle matching across more than two databases (which can be challenging using frequentist hypothesis testing approaches).
At a practical level, astronomers often misinterpret frequentist results; e.g., $p$-values from hypothesis tests are often interpreted as (Bayesian) probabilities for hypotheses of interest, or as conditional error rates, which are closely related to Bayesian model probabilities (Berger 2003).
We see this as indicating that Bayesian results more directly and intuitively answer the questions astronomers ask of their data than frequentist methods, at least for these problems.

\begin{figure}[t]
\centerline{\includegraphics[width=.7\textwidth]{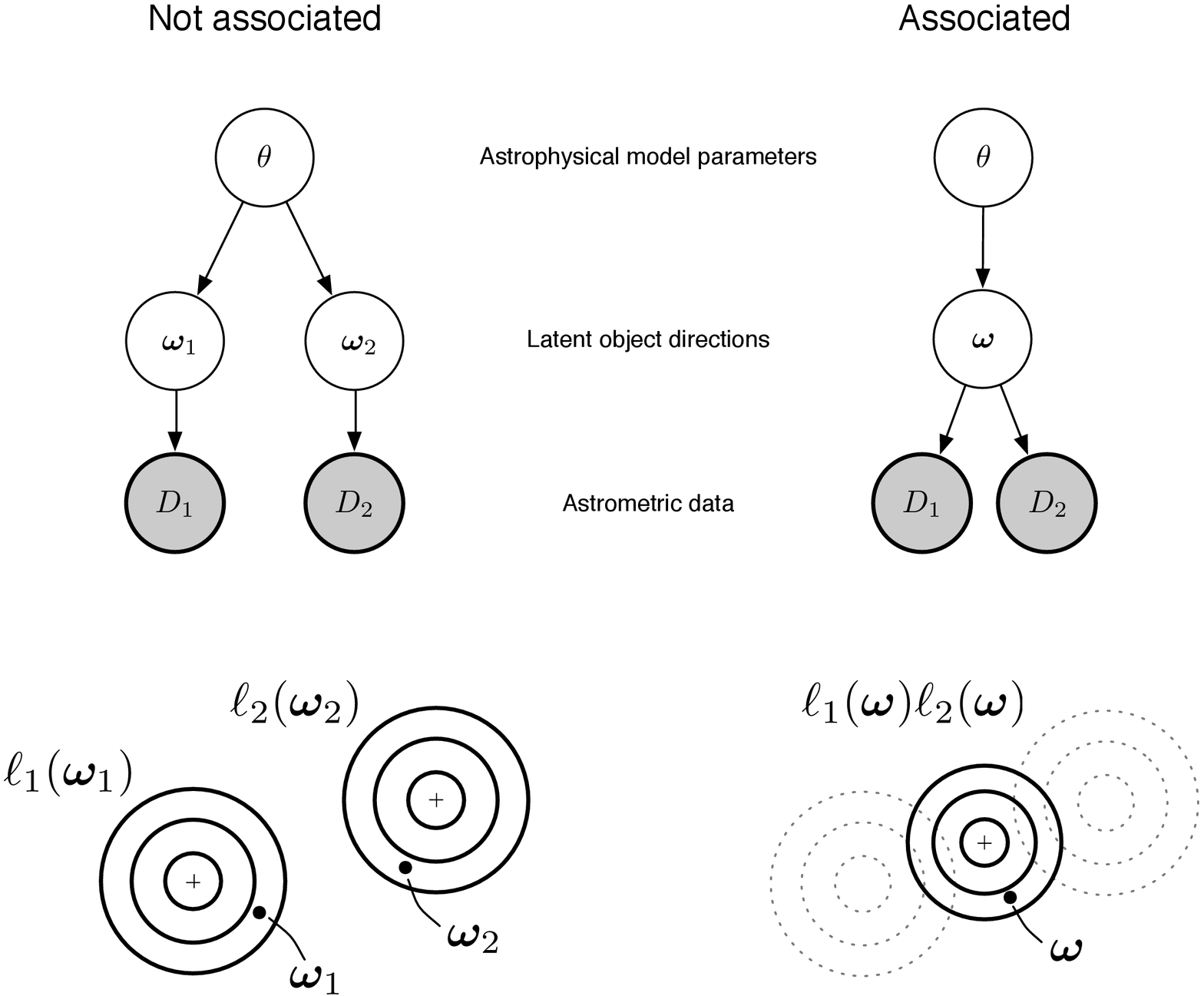}}
\caption{Multilevel (hierarchical) models for two sources.
{\em Left}: Model considering the sources to be observations of two separate objects with (latent) directions $\drxn_1$, $\drxn_2$; the source likelihood functions, $\ell_i(\drxn)$, are interpreted as describing independent uncertainties for each direction.
{\em Right}: Model considering the sources to be observations of the same underlying object with (latent) direction $\drxn$; the source likelihood functions describe independent measurements of the same direction, with their product describing the uncertainty of the combined data.}
\label{fig:PairMLMs}
\end{figure}

Hierarchical Bayesian cross-identification borrows ideas from several areas of applied statistics, although in some cases with a novel twist:
\begin{itemize}
\item {\em Partition models:}
Fundamentally, the approach frames cross-identification problems using partition models that calculate probabilities for different partitions of the data into singletons (single sources assigned to unique objects) and associated multiplets (sources that share a common object of origin, and thus a common latent direction).
LLW96 explicitly framed coincidence assessment in this way.
Hartigan (1990) developed a flexible theory for {\em product partition models} (PPMs), and the LLW96 approach can be seen as a PPM framework.
In a similar vein, Sadinle (2014) uses a Bayesian partition model to treat the problem of deduplication in homicide record systems.
\item {\em Graphical (hierarchical) models:}
The probabilistic framework is hierarchical, with separate levels describing measurement errors and other probabilistic ingredients.
\cfig{fig:PairMLMs} schematically depicts the hierarchical structure for the simple case of modeling data from two sources.
An upper level describes the distribution of directions across populations of objects, potentially with parameters, $\theta$, describing population properties such as the sky densities.
The lower level describes measurement error (and possibly selection effects in more complicated settings), providing the connection between latent object directions and the astrometric data used for direction measurements (e.g., image data).
The association and non-association models correspond to different graphs with different conditional independence structure; cross-matching can thus be seen as an example of graphical inference, with inference producing a probability distribution over possible graphs describing the data.
\item {\em Measurement error models:}
A primary aim of directional cross-identification is accounting for directional measurement error.
Our framework corresponds to a Bayesian measurement error model (Carroll, Ruppert, and Stefanski 2006, ch.~9), with the directional uncertainties accounted for in a manner similar to random effects models.
\item {\em Directional statistics:}
With directions on the sky as the fundamental observable, probabilistic cross-identification relies on distributions featured in directional statistics (Mardia 1972 and Fisher, Lewis and Embleton 1987).
In particular, the Fisher distribution takes the place of the normal distribution as a standard quantification of measurement uncertainty.
\item {\em Combinatorics for coincidences:}
Until recently, statistical approaches to studying coincidences in record linkage have been predominantly frequentist, relying largely on calculating probabilities for spurious coincidences under a null hypothesis of no true associations (e.g., Fellegi \& Sunter 1969; for recent generalizations, see Sadinle and Fienberg 2013).
In the broader context of coincidence assessment, this approach dates back at least to the famous birthday problem (see Diaconis and Mosteller 1989 for a review), and frequently involves combinatorial calculations, e.g., of the number of partitions of the data into spurious associations.
Although we adopt a Bayesian approach, similar calculations appear in assigning prior probabilities to candidate associations.
\item {\em Multiple testing:}
With two sources, there are two possible association hypotheses:  a no-association hypothesis, and a single (doublet) association hypothesis.
With three sources, besides the no-association hypothesis, there are three doublet-and-singlet association hypotheses, and one triplet association hypothesis.
The number of possible associations grows combinatorially with the size of the data.
Cross-identification thus involves comparing large numbers of hypotheses, the topic of much recent research on multiple testing.
Bayesian methods can naturally account for hypothesis multiplicity, with hierarchical models able to adapt multiplicity corrections to the data (Scott and Berger 2006).
\item {\em Clustering:}
Cross-identification appears similar in some respects to clustering (especially in the intra-catalog setting), in the sense that an association may be viewed as a cluster.
However, in contrast to unsupervised clustering, there is often important information available that should guide cluster assignments (i.e., associations).
Measurement error sets a scale for associations, but in a more complicated way than is typical in clustering when (as is typically the case) the errors are heteroskedastic, and sometimes of greatly different scale across catalogs.
More fundamentally, in contrast to clustering, there are natural latent variables for directional association, namely the true directions to the underlying objects.
\item {\em Mixture models:}
Mixture models are often used as generative models for clustering.
A skymap of associated sources resembles a scatterplot of samples from a mixture model, and some astronomical cross-identification problems have been framed as mixture model problems (e.g., Watson et al.\ 2011; see Sadinle and Fienberg 2013 for a non-astronomical example).
In the hierarchical Bayesian framework reviewed here, mixture models naturally arise in some regimes (e.g., for asymmetric cross-matching where many-to-one matches are permitted, as in the UHECR problem; Loredo 2012).
But the framework is more general, allowing models not easily expressible as mixture models.
\end{itemize}

\subsection{Partition models for directional cross-identification}

We now outline a general framework for hierarchical Bayesian directional cross-identification.
Let $D_{ic}$ denote the astrometric data for source (item) $i$ in catalog $c$, with $C$ catalogs each summarizing results from analyzing data from $N_c$ sources.
We interpret the catalog summary for source $i$ in catalog $c$ as describing a likelihood function, $\slike_{ic}(\drxn)$, for the uncertain direction, $\drxn$, a unit 3-vector pointing to the object producing the data for the source.
That is, $\slike_{ic}(\drxn) = p(D_{ic}|\drxn)$, the sampling distribution for the astrometric data (possibly marginalized with respect to other source parameters needed to model the data), considered as a function of $\drxn$.
For example, when the catalog reports a best-fit direction and an angular scale for azimuthally symmetric direction uncertainties, we may take the likelihood function to be (approximately) proportional to a Fisher distribution with location parameter given by the best-fit direction, and concentration parameter determined by the reported error scale (we describe this further below).

An association hypothesis corresponds to a partition of the data into subsets, with single-element subsets (singletons) corresponding to sources with unique objects, and multiple-element subsets (multiplets) corresponding to sources associated with the same underlying object.
One way to describe such a partition is by listing $(i,c)$ tuplets in subsets. 
E.g., for three catalogs with \mbox{$N_c\!=\!3$}, 2, and 1, the partition 
%
\begin{align}
\partn = \{\; &\{(2,1)\},\nonumber\\
    &\{(3,1),(1,2)\},\nonumber\\
    &\{(1,1), (2,2), (1,3)\}\; \}
\end{align}
denotes a hypothesis with one singlet (source 2 in catalog 1), one doublet (source 3 in catalog 1, and source 1 in catalog 2), and one triplet (sources 1, 2, and 1 in catalogs 1, 2 and 3).


Alternatively, we can index the $\Nobj$ objects underlying a partition (the number of subsets in the partition description) with an integer $o=1$ to $\Nobj$, and introduce source labels, $\slbl_{ic}$, that specify the object hypothesized to produce the data for source $i$ in catalog $c$.
We let $\sset_o$ denote the set of sources associated with object $o$, i.e., $\sset_o = \{(i,c): \slbl_{ic}=o\}$.
We let $m_o$ denote the multiplicity of the associations for object $o$ (1 for singlets, 2 for doublets, etc.), and $\cset(o)$ denote the list of catalogs containing the sources associated with object $o$.
We sometimes refer to $\cset(o)$ as the {\em association type} for object $o$.
For example, in an analysis of optical, infrared, and ultraviolet catalogs, there may be three different singlet association types (one for each catalog), three doublet association types (corresponding to the three possible pairs of catalogs), and one triplet association type (excluding associations that would include multiple sources from one catalog).
A complication of this object index description is that it suffers from a label-switching problem, i.e., sets of labels that correspond to permutations of the object indices correspond to the same association hypothesis.

A key property of association hypotheses is that the data comprising the associations in a partition are conditionally independent, given the partition.
That is, the likelihood function for a partition factors,
\be
\like(\partn)
 \;\equiv\; p(\{D_{ic}\}|\partn)\nonumber\\
 \;=\; \prod_o \mlike_o,
\label{like-partn}
\ee
where the marginal likelihood for the associations assigned to object $o$ is
\be
\mlike_o = \int d\drxn \, \rho_{\cset(o)}(\drxn) \prod_{(i,c)\in \sset(o)} \slike_{ic}(\drxn),
\label{obj-like}
\ee
where $\rho_{\cset}$ denotes the distribution of directions for objects producing sources in the catalog set $\cset$.

For comparison, the marginal likelihood for non-association of the sources in $\sset(o)$ is
\be
\namlike_o = \prod_{(i,c)\in \sset(o)} \int d\drxn \, \rho_{c}(\drxn) \, \slike_{ic}(\drxn),
\label{obj-nalike}
\ee
where $\rho_c$ denotes the distribution of directions for singletons in catalog $c$.
The ratio of the marginal likelihoods for the association and non-association hypotheses is the {\em Bayes factor} in favor of association,
\be
B_o \equiv \frac{\mlike_o}{\namlike_o}.
\label{B-obj}
\ee
The Bayes factor quantifies the evidence in the data favoring a particular association.
However, it ignores prior probabilities for the hypotheses, thereby not accounting for the multiplicity of association hypotheses and the prior probabilities for different types of associations (e.g., the relative sky densities of populations producing different types of associations, i.e., different catalog sets, $\cset(o)$).

In general, the prior probability distribution over partitions will depend on population parameters, $\pparams$, that may specify the relative densities and other properties of astrophysical populations producing different types of associations.
We make the parametric product partition model assumption that the conditional prior over partitions (conditioned on $\pparams$) has the product form,
\be
p(\partn|\pparams) = K \prod_o k(\sset_o, \pparams),
\label{PPM-prior}
\ee
where, in PPM terminology, $k(\sset_o, \pparams)$ is called the {\em prior cohesion} for the subset of sources associated with object $o$, and $K$ is a normalization constant.
The prior cohesions may be specified uniquely only up to a set of proportionality constants that merely redefine $K$ (Hartigan 1990).
They can account for the relative numbers of different types of a priori associations.
Different types of cross-identification problems---such as intra-catalog associations, asymmetric, many-to-one host-target associations, and symmetric one-to-one associations---will have different choices of prior cohesions.

A key property of PPMs is that the factorization properties of the likelihood and prior ensure that the conditional posterior over partitions, $p(\partn|\pparams,\{D_{ic}\})$, also factors into a product of posterior cohesions, proportional to the product of the prior cohesions and the marginal likelihoods for the objects comprising a partition.
The conditional posterior can be used to infer the population parameters by introducing a population parameter prior, $p(\pparams)$, and calculating the the marginal distribution for $\pparams$,
\be
p(\pparams|\{D_{ic}\})
  \propto p(\pparams) \sum_\partn p(\partn|\pparams,\{D_{ic}\}).
\label{pparam-post}
\ee
The UHECR calculation briefly described above proceeded this way, using Markov chain Monte Carlo methods to marginalize over many-to-one UHECR-AGN association hypotheses.
If instead one is interested in assessing candidate associations, formally one should marginalize over $\pparams$ to find a marginal posterior over partitions,
\be
p(\partn|\{D_{ic}\})
  \propto \int d\pparams\, p(\pparams) p(\partn|\pparams,\{D_{ic}\}).
\label{partn-post}
\ee
In practice, the marginal distribution may sometimes be accurately approximated by conditioning on maximum likelihood estimates for $\pparams$, producing the plug-in partition posterior, $p(\partn|\hat\pparams, \{D_{ic}\})$, where $\hat\pparams$ denotes the maximum likelihood population parameter estimate.
This approximation is essentially a type of empirical Bayes procedure (i.e., ignoring the uncertainty in $\pparams$ that would be accounted for in a fully hierarchical Bayesian treatment); we describe such a calculation next.

\subsection{Symmetric two-catalog cross-identification}

As a simple but important example, consider symmetric cross-identification of two catalogs, with the goal of finding secure associations between the catalogs using the plug-in partition posterior.
Let $I_1(o)$ and $I_2(o)$ denote the $(i,c)$ identification of the sources associated with doublet object $o$.
The Bayes factor for a {\em particular} candidate association is the ratio of a doublet marginal likelihood to a two-singlet non-association marginal likelihood,
\be
B_o
  \;=\; \frac{\mlike_o}{\namlike_o}\nonumber\\
  \;=\; \frac{\int\!d\drxn \, \rho(\drxn) \slike_{I_1}(\drxn) \slike_{I_2}(\drxn)}
{\int\!d\drxn_1 \, \rho(\drxn_1) \slike_{I_1}(\drxn_1) \  \int\!d\drxn_2 \, \rho(\drxn_2) \slike_{I_2}(\drxn_2)},
\label{B-2cat}
\ee
where we have assumed the direction distributions for singlets and doublets are the same, $\rho(\drxn)$, e.g., $\rho(\drxn) = 1/(4\pi)$ for isotropic distributions.
For a simple population model, the prior cohesions will lead to a prior probability for a doublet, $\beta$ (with $\beta$ specified by $\pparams$), with the prior probability for non-association of the doublet members equal to $1-\beta$.
The conditional posterior odds for the association is the product of the prior odds, $\beta/(1-\beta)$, and the Bayes factor;
\be
O_o(\beta) = \frac{\beta}{1-\beta} B_o.
\label{odds-beta}
\ee
The conditional posterior probability for the association can be written in terms of these quantities as
\be
P_o(\beta)
  = \frac{\beta \mlike_o}{(1-\beta)\namlike_o + \beta\mlike_o}
  = \frac{O_o(\beta)}{1 + O_o(\beta)}.
\label{P-beta}
\ee

To estimate $\beta$, we must pool information from all possible associations, via the conditional posterior over partitions.
For the plug-in approximation, we ignore the prior for $\pparams$ and focus on the likelihood function for $\beta$, which is found by summing over partitions, as in \ceqn{pparam-post}.
If we permit one-to-many matches, which is a reasonable assumption in astronomy at least for resolved sources, whose images at different wavelengths might be segmented differently, the resulting sum of products of marginal likelihoods can be rewritten as product of sums (Loredo 2012; an analogous result is well known for mixture models, see Bernardo and Gir\'on 1988),
\be
\like(\beta) \propto \prod_o \left[ (1-\beta)\namlike_o + \beta\mlike_o\right].
\label{like-beta}
\ee
%
%
The maximum likelihood estimate $\hat{\beta}$ can be obtained analytically.
By setting the derivative of the log likelihood to zero, we get
\begin{equation}
\sum_{o} \hat{P}_{o} = \hat{\beta} \sum_{o}1
\end{equation}
where $\hat{P}_o$ is the posterior at $\hat{\beta}$ and the sum of 1 is the total number of possible assocations, equal to the product of the number of sources in the analyzed catalogs, i.e., \mbox{$N_1 N_2$} in our 2-way matching scenario.
The expected number of matches is then calculated as
\begin{equation}
N_{\star} = \hat{\beta} N_1 N_2.
\end{equation}
If the prior over $\beta$ is not highly informative, the posterior is approximately Gaussian with mean $\hat{\beta}$ and its standard deviation may be obtained by taking the second derivative of the log likelihood, giving
\begin{equation}
\frac{1}{\sigma_{\beta}^2} =
\frac{\sum 1}{\hat{\beta}^2(1\!-\!\hat{\beta})^2}
\left[
\frac{\sum\hat{P}_{o}^2}{\sum 1} 
- \left(\frac{\sum\hat{P}_{o}}{\sum 1}\right)^2
\right]
\end{equation}
where all sums go over $o$. 
It is illuminating to evaluate this formula in two limiting regimes.
For very large uncertainties, the observations do not provide any constraints. The likelihood functions are constant, \mbox{$\slike_{I}(\drxn)\rightarrow{}A$}, and \mbox{$\mlike_o=\namlike_o=A^2$}, which yields \mbox{$B_{o}=1$} as expected. The likelihood $\like(\beta)$ is flat and there is no unique $\hat{\beta}$ maximum.
In the other extreme, when the uncertainties are very small compared to the typical separations between objects, the posteriors $P_{o}$ are approximately 0 or 1. The number of expected matches becomes
\begin{equation}
N_{\star} = \sum_{\rm{}all} P_{o} 
\approx\!\!\!\!\sum_{\rm{}matches}\!\!\!\!\!\!P_{o} 
\approx\!\!\!\!\sum_{\rm{}matches}\!\!\!\!\!\!1 
\equiv N_+
\end{equation}
hence
\begin{equation}
\hat{\beta} \approx \frac{N_+}{N_1 N_2},
\end{equation}
and the variance is \mbox{$\sigma_{\beta}^2\rightarrow{}{\hat{\beta}}$}.
The larger the catalogs, the smaller $\hat{\beta}$ becomes, hence an empirical Bayes approach is well justified in this regime, which holds in many applications.

\section{Applications in astronomy}
\label{sec:app}

Cross-identification of objects across separate catalogs is at the heart of all multi-color and time-domain (longitudinal) studies in astronomy. The probabilistic approach outlined in Section~\ref{sec:bayes} has been recently applied in and extended to a number of scenarios.
In many commonly arising cases, the new method has no extra cost as it is analytically tractable, but it can accommodate more complicated problems for which no other adequate solution had been proposed before.

\subsection{Closed-form Bayes factor for directional data}
\label{sec:analytic}

The normal distribution emerges often in descriptions of uncertainties in observational data in astronomy.
It is sometimes justified where a large number of additive effects play roles in shaping up the probability, via the central limit theorem (CLT). 
In other cases, it may be justified via Taylor expansion of the logarithm of the sampling density or likelihood function (in the manner of the Laplace approximation), or via appeals to sufficiency or maximum entropy arguments.
One of the earliest application of Gauss's bell curve was in fact in astronomy (Gauss 1809), and even today the sampling distribution for most catalog errors is often approximated by the normal distribution.
Catalogs will report, for instance, that the ``$1\sigma$'' directional uncertainty is 0.1~arc seconds (an unfortunately ambiguous designation; see below).
But for closed topological manifolds such as a (one-dimensional) circle or the (two-dimensional) surface of the celestial sphere, there is no one distribution sharing all the appealing properties of the normal distribution, and some care must be taken in calculating and reporting directional uncertainties.
For example, on the circle, the circular CLT leads to a wrapped normal, while maximum entropy or moment-based sufficiency arguments lead to the von Mises distribution.

On the sphere, a simple directional distribution used to describe azimuthally symmetric uncertainties is the Fisher (or von Mises-Fisher) distribution  (Fisher 1953), sometimes known as ``the Gaussian on the sphere.''
For a sampled direction denoted by a unit vector $x$, the Fisher probability density is
%
%
%
\begin{equation}
f(x;\omega,\kappa) = \frac{\kappa}{4\pi\,\sinh{}\kappa}\
           \exp \Big( \kappa\,\omega\!\cdot{}\!x \Big),
\end{equation}
with two parameters, a location parameter,  $\omega$ (a unit vector specifying the direction of the mode), and a concentration parameter, $\kappa$.
Here $\omega\!\cdot{}\!x$ denotes the vector dot product of the two directions, equal to the cosine of the great circle angle between them.
The Fisher distribution can be used to characterize directional uncertainty in astronomy observations.
In such cases, analysis of raw image data produces a likelihood function for the source direction, $\omega$, that may be well approximated by a Fisher distribution with best-fit direction $x$ and concentration $\kappa$ (both of them functions of the image data).
It is particularly useful when the uncertainty is large, in which case a bivariate normal distribution on the tangent plane will poorly represent uncertainties.
When the concentration parameter \mbox{$\kappa\!\rightarrow\!0$}, the distribution becomes uniform on the surface of the unit sphere, \mbox{$f(x)\!\rightarrow\!1\big/4\pi$}, and the observations provide no constraints.
In the limit of high precision---large $\kappa$---the Fisher distribution corresponds to a symmetric bivariate normal distribution on the tangent plane with \mbox{$\kappa\!=\!1/\sigma^2$}, where $\sigma$ is the (marginal) standard deviation in radians for a single coordinate.%
\footnote{More precisely, in the $\kappa\gg 1$ limit, the Fisher density becomes an uncorrelated bivariate normal with respect to locally cartesian arc length coordinates about the mode on the unit sphere.
The standard deviation in each of the coordinate directions is $\sigma \approx 1/\kappa^{1/2} \approx 57.3^\circ/\kappa^{1/2}$ in this limit.
The angular radius containing 68.3\% probability may be found by integrating the Fisher density function.
For $\kappa\gg 1$ the angle containing probability $P$ is $\theta_P \approx -(2/\kappa)\log(1-P)$.
For $P=0.683$, we have $\theta_P \approx 2.30/\kappa$ radians, or $\theta_P \approx 86.9^\circ/\kappa^{1/2}$.
The ``$1\sigma$ radius'' reported in a catalog may denote either a coordinate standard deviation, or an error circle radius; since these angular scales are different, care must be taken in interpreting reported uncertainties.}


For both the Gaussian and Fisher distributions, the marginal likelihoods and the Bayes factors can be calculated analytically (LLW96; Graziani \& Lamb 1996; BS08). 
Assuming a uniform prior density on the entire sky,
the more general Fisher distribution yields a simple closed-form expression for the Bayes factor:
\begin{equation} \label{eq:bfsinh}
B=\frac{\sinh{}\kappa}{\kappa}\,\prod_{k=1}^{K} \frac{\kappa_k}{\sinh{\kappa_k}}, 
\ \ \ \ \ \ {\rm{with}} \ \ \ \ \ \ 
\kappa = \left| \sum_{k=1}^{K} \kappa_k x_k \right|.
\end{equation}
If all positional measurements are highly accurate \mbox{($\kappa_k\!\gg{}1$)}, we get back a more intuitively accessible expression corresponding to the Gaussian limit:
\begin{equation} \label{eq:bfprec}
B = 2^{n-1} \frac{\prod \kappa_k}{\sum \kappa_k} \exp \left\{ -
  \frac{\sum_{k<l} \kappa_k \kappa_l \varphi_{kl}^2}{2\sum{\kappa_k}} \right\},
\end{equation}
where $\varphi_{kl}$ is the angle between $\vx_k$ and $\vx_l$ measured directions.
In the 2-way case, the dimensionless Bayes factor simplifies to
\begin{equation} \label{eq:bf2}
B = \frac{2}{\sigma^2_1 + \sigma^2_2}
         \exp \left\{-\frac{\varphi^2}{2(\sigma^2_1 + \sigma^2_2)} \right\},
\end{equation}
where all quantities are in radians, and \mbox{$\sigma_k^2=1/\kappa_k$} as before.
With these equations in hand, one can directly evaluate a reliable quality measure for each candidate match---no fundamental change is required in processing pipelines using more conventional metrics and thresholds, and practically no extract computational cost is incurred.
Note that for equal accuracies (i.e., homoskedastic measurement error), a cut on the Bayes factor \mbox{$B=B(\varphi;\sigma_1,\sigma_2)$} is equivalent to thresholding the angular separation as $B$ only depends on $\varphi$.

\begin{figure}[t]
\centerline{\includegraphics[width=.75\textwidth]{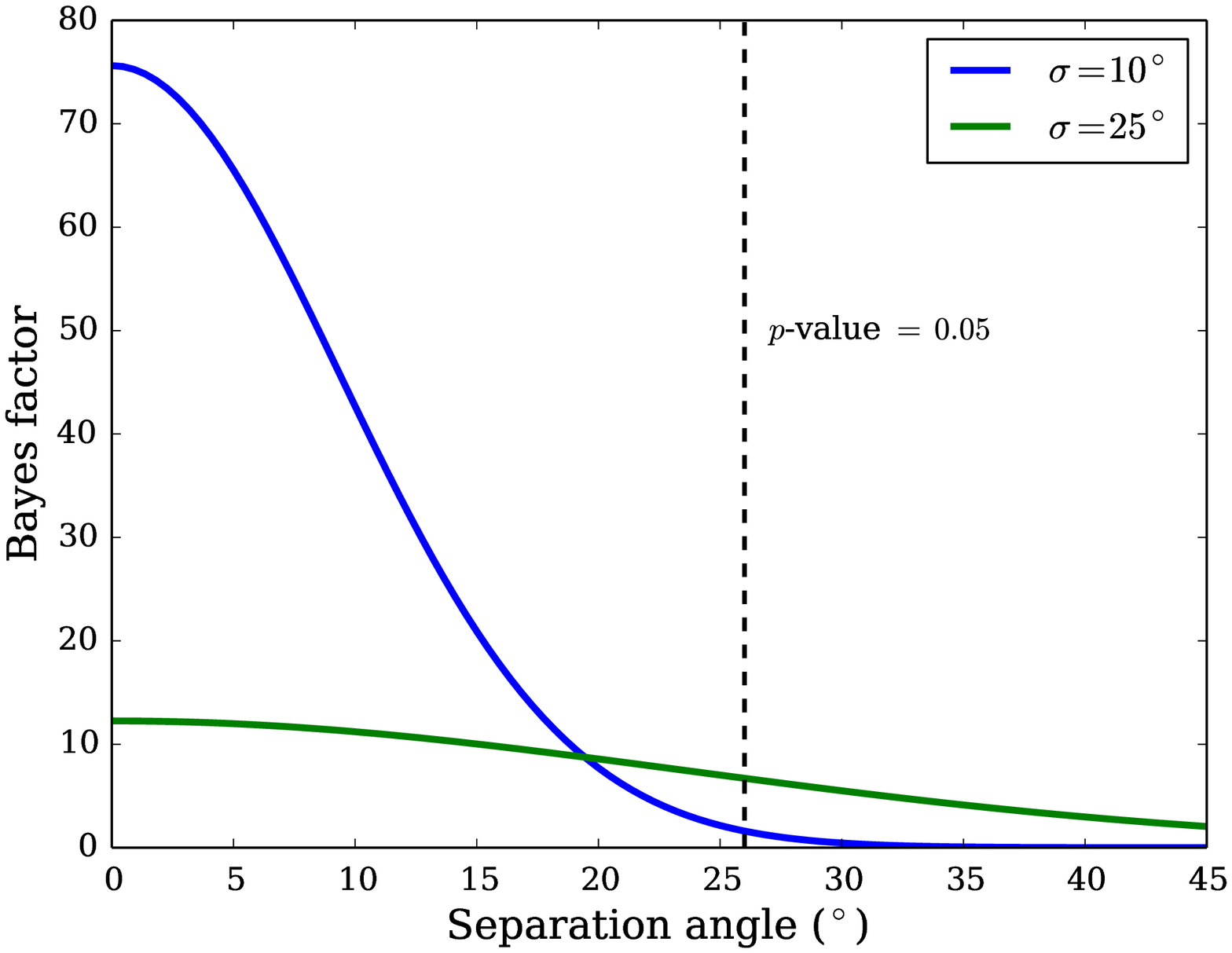}}
\smallskip
\caption{Doublet Bayes factor favoring association of two sources as a function of angular separation, for sources with the same error circle radii, for two cases: $\sigma=10^\circ$ (blue) and $\sigma=25^\circ$ (green).
Dashed vertical line shows the angular separation corresponding to a $p$-value of 0.05 under the null hypothesis of no association and two isotropically distributed sources.}
\label{fig:BF}
\end{figure}

\cfig{fig:BF}\ plots the Bayes factor for a doublet association as a function of the angular separation of the sources, for sources with equal angular uncertainties (i.e., \ceqn{eq:bfsinh} with $K=2$ and $\kappa_1 = \kappa_2$).
The two curves correspond to 68.3\% error circle radius values of $\sigma=10^\circ$ (blue) and $25^\circ$ (green); these are scales typical of GRB direction uncertainties (and much larger than the arcsecond scales of stellar direction uncertainties).
The vertical dashed line shows the angle ($\approx 26^\circ$) corresponding to a $p$-value of 5\% under the null hypothesis of no association.
Notably, the $p$-value does not depend on $\sigma$ because the convolution of the error distribution and the uniform sky density remains uniform (of course, a careful hypothesis testing approach would also consider the power against alternatives, which would depend on $\sigma$).
The Bayes factor increases with decreasing separation, with the peak value (at zero separation) depending on the direction uncertainty in a sensible way.
The peak Bayes factor for $\sigma=25^\circ$ is only $\approx 12$, which would indicate positive but not entirely compelling evidence for association.
Although zero separation is an unlikely outcome under the null (with $p$-value equal to zero independent of the value of $\sigma$), when $\sigma$ is large, exact coincidence ($\varphi = 0$) does not have a significantly larger probability density than nonzero separations, moderating the weight of evidence for association.
In this way, the Bayes factor nicely captures the capability of the observations to distinguish the hypotheses; for large error scales, no data will provide compelling evidence for association.
Note also that the Bayes factor at $26^\circ$ separation is \emph{smaller} for $\sigma=25^\circ$ than for $\sigma=10^\circ$.
For the smaller error scale, separations near $26^\circ$ are very unlikely if the sources are associated, so the association hypothesis is penalized more strongly than is the case when the error scale is large.

\subsection{Limited sky coverage}
\label{sec:footprint}

The available sky is limited in each experiment by the geometry and movement of the instrument and the Earth. The location of each telescope determines the possible directions observations can be taken. Astronomers are painfully aware of these boundaries and optimize the surveys accordingly. The actual sky coverage is the union of all observations, which are recorded along with summary catalogs. These are either captured in (hierarchical) pixel maps (e.g., HEALPix, Igloo, SDSSpix and HTM) or are described analytically by equations (e.g., Budav\'ari, Szalay, Fekete 2010).  
On top of these representations specialized search engines are built. A successful example is the Virtual Observatory Footprint Service at \url{http://voservices.net/footprint} that hosts a large number of coverage maps.

The geometric constraints can be used to refine the isotropic directional prior, which in the (usual) case of large concentration parameters becomes approximately constant over the area of the sky coverage $\Omega$ for all detections within and away from the boundary.
The new prior \mbox{$\rho'(\drxn)=1/\Omega$} enters the equations in two places. 
New variants of the marginal likelihoods, denoted by the primed symbols, e.g., $\mlike_o'$, will be larger because
\mbox{$\mlike_o/\mlike_{o}'=(\Omega\big/4\pi)\leq1$}
and similary \mbox{$(\Omega\big/4\pi)^K$}, hence the all-sky Bayes factor can be expressed with the limited-sky variant as
\begin{equation}
B_{o} = \left(\frac{\Omega}{4\pi}\right)^{\!1\!-\!K}\!B'_{o}.
\end{equation}
The constraints on $\beta$ also change and the MLE becomes
\mbox{$\hat{\beta}' = N_{\star}' \Big/ \prod N_k'$}, where the new values are the scaled down numbers for the given area, so that
\mbox{$N'=(\Omega\big/4\pi) N $}. 
With this the relation to the all-sky estimate becomes
\begin{equation}
\hat{\beta} = \left(\frac{\Omega}{4\pi}\right)^{\!K\!-\!1} \hat{\beta}' .
\end{equation}
Considering that for small $\beta$, the posterior practically only depends on the product of $\beta$ and $B_o$, the result 
\begin{equation}
P_{o} \approx \frac{\beta{}B_o}{1+\beta{}B_o}
\end{equation}
is not sensitive to picking the all-sky vs.\ the limited-sky equations. In other words,
the posterior probability really only depends on the local (surface) density of sources and not the total number, which is very intuitive and advantageous for situations where $\beta$ varies on the sky. For example, there are many more stars in the directions along the galactic plane of our disk-shaped Galaxy than looking perpendicularly out of the Milky Way.

\subsection{Stars with unknown motion} 
\label{sec:pm}

Stars pose a significant challenge for cross-identification methods because they move through space with sufficient speed to alter their apparent positions measurably with time. 
The perpendicular component of the movement to the line of sight direction (i.e., the component in the plane of the sky) is called {\em{}proper motion} and it varies strongly depending on the distance of the star and its cosmic history.
The analytic equations developed for matching static objects is clearly not adequate when the stars move between the observations by a significant amount compared to the astrometric accuracy of the measurements.

Kerekes et~al.~(2010) extended the probabilistic approach method to include unknown proper motion in the hypotheses.
To evaluate the marginal likelihoods one has to model the directions of objects as a function of time. The new parameter is the proper motion $\vmu$ and in addition to the positional measurements one also has to keep track of time. If $\omega$ is the true direction of an object at our reference epoch, the probability density of finding it at some $\vx$ will also depend on the time difference $\Delta{}t$. Formally we can write the likelihoods of the hypotheses in the Bayes factor as before but now using all the parameters of the new model. For example, the likelihood for the match becomes
\begin{eqnarray} \label{eq:bfmu}
\mlike_{o} = \int\!\!d\drxn\int\!\!d\mu\ \rho(\drxn,\mu)\,L_{o}(\drxn,\mu),
\end{eqnarray}
where the likelihood function $L_{o}(\cdot)$ compares the detected directions to the appropriate model directions at the right time. In this simple model of small movements on the sky, the  transformation of the directions is 
\begin{equation} \label{eq:mu}
\omega' = \frac{\omega + \vmu\Delta{}t}{\left| \omega + \vmu\Delta{}t \right|},
\end{equation}
which one can directly use in, for example, the Fisher distribution.

The integrals can only be evaluated with a known prior in hand, which has now two variables. The joint density can be written as
\mbox{$\rho(\omega,\vmu)\!=\!\rho(\omega)\,\rho(\vmu|\omega)$}
where the first term is the directional prior, same as before, and the second describes the possible proper motions (amplitude and orientation) as a function of the location on the sky.
The simplest assumption is a uniform distribution up the a maximum of $\mu_{\rm{max}}$ independently from the position,
\begin{equation} \label{eq:uconst}
\rho(\vmu|\omega) =  \left\{\begin{array}{c l}
           1 \big/ \pi\mu_{\rm{max}}^{2}  & \quad \mbox{if\ $|\vmu|<\,\mu_{\rm{max}}$}\\
           0 & \quad \mbox{otherwise.}\\ \end{array} \right.
\end{equation}
Alternatively one can also use observations to build up more realistic priors that depend on the position or other measured properties of the stars.

\begin{figure}
\centerline{\includegraphics[width=350pt,height=170pt]{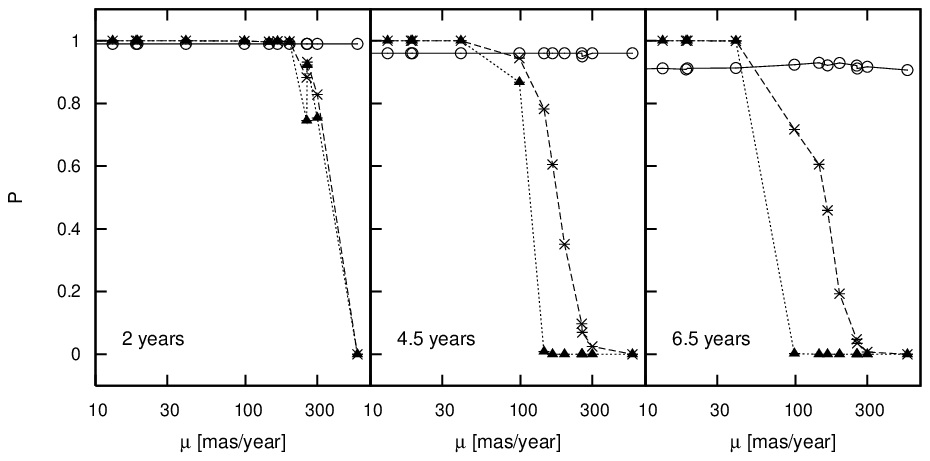}}
\caption{Probability of 2-way associations as a function of proper motion. As we look at longer time differences between the epochs, our models yield increasing different answers.}
\label{fig:pm1}
\end{figure}

\begin{figure}
\centerline{\includegraphics[width=350pt,height=170pt]{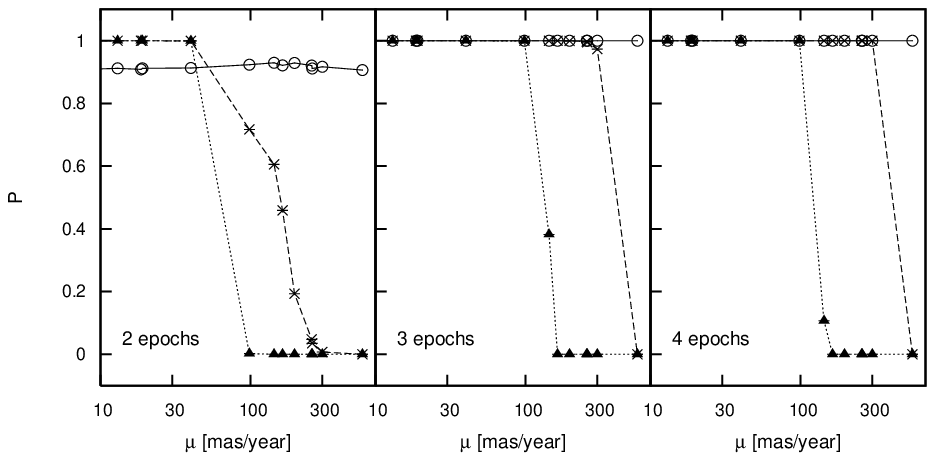}}
\caption{Probability of 2-, 3-, and 4-way star associations as a function of proper motion.
Introducing a 3rd intermediate epoch improves the cross-identification significantly.
The effect of adding a 4th measurement is not so dramatic.}
\label{fig:pm2}
\end{figure}

Kerekes et~al.(2010) used a sample of stars with known proper motions to demonstrate the power of the method as a function of increasing time-difference and angular separation as well as the number of available exposures. 
In Figure\,\ref{fig:pm1} we see the 2-way matching results for varying separations in time. As we increase the time baseline (from left to right) the static model (triangles) starts to reject the faster stars but the proper motion models assign finite probabilities, even if not too large. The constant prior yields a constant posterior (open circles) that becomes lower with time.
Perhaps even more interesting is Figure\,\ref{fig:pm2} where the same time baseline is shown but with the inclusion of intermediate epochs. The previous 2-epoch results jump to high probabilities for 3- and 4-epoch observations. While their superior quality might be somewhat surprising at first, it is actually easy to understand: Proper motion is well-approximated by a movement along a great circle. 
The big difference between two and more epochs is the fact that it is always possible to precisely fit a great circle to two points on the sky but not to three or more, hence such configurations are very much rewarded.

%

\subsection{Beyond directional data} 
\label{sec:phot}

When the directional evidence is not conclusive, the same probabilistic approach can be applied to other measurements to improve our ability to cross-identify sources.
A natural extension is to use the measured source fluxes (flux is apparent brightness, in units of photons or energy per unit area per unit time), as those measurements are always available in experiments using imaging cameras. 
For example, for an object emitting with constant luminosity, we expect repeated measurements to be consistent with a constant flux, so we should seek to detect coincidence, not just in direction, but also in flux.

A number of semi-analytic and empirical models exist to simulate fluxes that can be used to calculate marginal likelihoods.
One challenge is that the prior distribution to use for fluxes is often unknown; in fact, the scientific questions motivating cross-identification tend to be exactly about cosmic demographics that such priors would encode. 
A generative hierarchical model can naturally account for these complications. 
Fundamentally, the statistical cross-identification problem cannot be discussed separately from this kind of physical modeling and classification.
Probabilistic cross-identification and astronomy research questions are intertwined and can be optimally addressed only in a joint analysis.
Astronomers are only just beginning to undertake such analyses (the UHECR problem of Section~\ref{sec:bayes} is an example).

Another challenge is that the source measurements we would like to use for cross-identification are often not independent. 
For example, measurement of the direction and the flux both rely on the same raw image data (counts of photons in pixels). 
The flux may be estimated using the sum of counts in all pixels that belong to the source, and the direction may be estimated from the average of the pixel positions using the same pixel counts as weights.
One can show that for objects with point symmetry, which is a good approximation for many sources in astronomy such as stars or galaxies, the total flux and the measured direction are uncorrelated.
But even in this symmetric case they are not independent (e.g., the estimated flux decreases as the direction parameter is moved away from the best-fit direction). 
A complete end-to-end simulation of observations of model galaxies that could account for this is, however, computationally very expensive. 
Considering that today's surveys have seen many hundreds of millions of sources in hundreds of terabytes of images, a rigorous and thorough statistical approach is very challenging.

Marquez et~al.\ (2013) considered a toy model where the flux and directional likelihood functions are considered independent and implemented the numerical integrals to assess the feasibility of such studies based on currently available models and tools.
Even this naive model was found to improve the matching on real observations where multiple matches were equally possible based purely on the directional data.
The fact that galaxies do not have arbitrary colors helps tremendously to reject bad candidate associations.

\section{Practical Considerations}
\label{sec:prac}

In practice a significant computational effort is required even for the simplest static matching scenario to quickly identify possible candidates and reject the obviously bad ones. 
The good news is that for many large collections of detections, direction errors are typically small and the finite area of the surface of the sphere (and the survey footprint) limits how many detections one can make. 
There have been several efforts undertaken to make cross-identification go fast in this regime, ranging from algorithm development to design of optimized hardware configurations.
%

\subsection{Recursive evaluation}

The total number of combinations in $K$-way matching is potentially enormous. Clearly one does not need to calculate the Bayes factor for all of them. If two observations are very far away, they can never be part of any probable association. 
This is the motivation behind a recursive approach, where collections of data are added to the association calculation iteratively. In every step the list of partial matches is pruned significantly.
Using the Fisher distribution in the limit of high accuracies, eq.(\ref{eq:bfprec}) can be rewritten to be evaluated incrementally with good approximation as
\begin{equation} \label{eq:bfinc}
\ln B = \ln \left( 2^{n-1} \frac{\prod \kappa_k}{\sum \kappa_k} \right) - \frac{1}{2}\sum_{k=2}^K \frac{a_{k-1}}{a_k} \kappa_k \vDel_k^2,
\end{equation}
with
\begin{equation}
a_k  =  \sum_{l=1}^k \kappa_l \ \ \ \ \ {\rm{and}} \ \ \ \ \
\vDel_k  =  \vx_k\!-\!c_{k-1},
\end{equation}
where $c_{k}$ is the unit vector of the best direction for the current
partial match,
\begin{equation}
c_{k} = \left( \displaystyle\sum_{l=1}^k \kappa_l \vx_l \right) \Big/ 
\, \left| \displaystyle\sum_{l=1}^k \kappa_l \vx_l \right|.
\end{equation}
In the $k$th step, the maximum search radius  
is computed from eq.(\ref{eq:bfinc}) such that it reaches a given Bayes factor threshold. In the calculation one can assume optimal subsequent matches with \mbox{$\Delta_k^2\!=\!0$} contributions.
We assign every source within that radius to each $k$-tuple sub-match, and go
to the next catalog.
From catalog to catalog we propagate the quantities that are
necessary to calculate the Bayes factor. The recursion formulas are
given by the following expressions:
\begin{eqnarray}
a_{k} &=& a_{k-1} + \kappa_{k}, \nonumber\\
q_{k} &=& q_{k-1} + \frac{a_{k-1}}{a_{k}} \kappa_k \Delta_{k}^2, \\
c_{k} &=& \left({c_{k-1} + \frac{\kappa_k}{a_{k}} \vDel_{k}}\right)
            \Big/\ {\left|\,c_{k-1} + \frac{\kappa_k}{a_{k}} \vDel_{k} \right|\nonumber}.
\end{eqnarray}
The initialization is $a_1\!=\!\kappa_1$, $q_1\!=\!0$ and $c_1\!=\!\vx_1$. 
This iterative approach enables the development of fast cross-matching tools.

\subsection{Online Service: SkyQuery}
\label{sec:skyquery}

Over the last two decades the way astronomy is done has changed significantly. Survey projects run dedicated telescopes, whose observations are automatically pipelined into science archives. Each project would typically operate its own database that was accessed differently. Cross-identification in particular became extremely difficult due to the large volumes of data and the fact that the data sets resided in different cities, countries or even continents. 
Under the umbrella of the International Virtual Observatory Alliance (IVOA; \url{http://ivoa.net}) there has been a significant effort to create standardized access protocols and data services. Among the first prototypes was a cross-match service called SkyQuery (Budav\'ari et al.\ 2003).
Given the large number of data sets it is impossible to create a master catalog of all data because of the combinatorial explosion in the required storage. 
SkyQuery was built to dynamically federate separate archives by performing cross-matching on the fly. It would match only the requested catalogs and filter the results to speed up the execution.
The first prototype used a likelihood ratio criterion and quickly became very popular. The second reincarnation, called Open SkyQuery, has been one of the major flagships of the IVOA for almost a decade. Due to high demands of the service, certain restrictions had to be imposed on the result sets, which did not discourage the users. 
The service was recently superseeded by the third generation engine that removed the previous limitations by using an inherently parallel infrastruture where the requests run on a cluster of machines (Budav\'ari, Dobos \& Szalay 2013). 

The flexibility of SkyQuery comes from the fact that it uses (an extended subset of) the Structured Query Language (SQL), with which many astronomers are already familiar because most archives support SQL requests. 
It also uses a better matching algorithm based on the directional Bayes factors.
In fact the extensions were introduced only to explicitly define a cross-identfication problem across multiple catalogs, while also capturing all relevant results from the statistical procedures. The matching algorithm is able to work with heteroskedastic directional uncertainties and custom matching thresholds but it also returns a number of quantities including, for example, the log Bayes factor and the best direction for each association. This is achieve by a construct where the algorithm selection and problem definition also pretends to be a table, whose results are available for the user. The service has been recently released for testing at \url{http://voservices.net/skyquery}.

\subsection{Graphics Processors} 
\label{sec:gpu}

Computer architectures evolve quickly. Their processing power has been increasing exponentially for the last half century. 
Scientists and programmers are used to changes that simply make their codes run faster but this might prove to be different in the near future.
Following Moore's law the number of transistors on a single chip, such as our microprocessors, is still growing today. The design patterns, however, have  dramatically changed recently. The reason is power consumption.
Individual processor cores are not speeding up any more; if anything their clock rates are being lowered to dissipate less heat.
If applications are to gain performance, they need to make use of new multicore technologies. 

The most promising parallel technology today originated in video processing. The graphical processing units (GPUs) on the most recent video cards are capable of not only rendering millions of pixels on our screens in a fraction of a second, but also of performing general-purpose computations. The GPU architecture is significantly different from CPUs. In many ways it is much simpler, which allows for new opportunities in taking the multicore paradigm to its extreme. A modern GPU can have 1536 cores that run close to 75,000 parallel threads at any given moment.

Lee \& Budav\'ari (2013) used C++ and NVIDIA's \mbox{C for CUDA} (Compute Unified Device Architecture) programming languages to build a scalable solution for cross-identification problems. 
Their cross-platform matching tool is implemented on multiple GPUs driven by separate worker threads on the CPU. 
Custom CUDA kernels find the matches by looking through a large number of candidate associations. The implementation is admittedly not trivial because finding the matches is not enough: returning them efficiently is an even bigger challenge on such a parallel architecture.
The extra implementations complexity, however, is worth the effort. The speedup is tremendous. For example, previously two of the largest astronomy surveys, SDSS and GALEX, with cardinalities of 350 and 150 million sources, respectively, could be cross-matched in an hour; the same is now possible in just a few minutes on a single machine.

\section{Outlook}
\label{sec:sum}

We have discussed cross-identification based on inputs from \emph{calibrated} astrometric catalogs (i.e., with celestial coordinate estimates and uncertainties for all sources).
Images produced by professional astronomers are calibrated, but images from amateur astronomers typically are not (e.g., only an approximate overall pointing direction for an image may be known; pixels will not be assigned accurate celestial coordinates).
For time-domain astronomy, particularly for detecting and studying transient sources, observations from amateur astronomers can provide crucial information.
One way to exploit such data is to locate significant sources in an uncalibrated image, and perform automated calibration based on matching the spatial pattern of sources to patterns of already known objects in trusted catalogs.
This is the kind of cross-matching a human observer performs in recognizing asterisms and constellations.
The \emph{Astrometry.net} service (Lang et al.\ 2010) performs this task, combining machine learning and statistical elements to find the best pointing, scale, and rotation of an optical or near-ultraviolet image, producing estimated coordinates for most sources in the image (but without calibrated uncertainties).%
\footnote{See \url{http://astrometry.net/} for the service and related publications.}
The underlying shape analysis problem is finding the alignment of point sets related by an uncertain geometric transformation, a problem that also arises in protein bioinformatics and other fields.
Several teams have developed flexible hierarchical Bayesian methods for such problems (e.g., Green and Mardia 2006; Kenobi and Dryden 2014).
It seems likely that statistical and computational methods could be profitably shared between these astronomical and bioinformatic applications.

An important issue for accurate statistical cross-identification is \emph{accurate representation of source direction uncertainties}.
For relatively bright sources, observed with instruments with an approximately axisymmetric angular response---conditions that hold for many optical and infrared survey catalogs---the axisymmetry and Gaussian-like falloff the the Fisher distribution may accurately describe the uncertainties.
But for dim sources (for which Gaussian-like behavior may be inadequate), and for instruments with complicated angular response, we need flexible alternatives to the Fisher distribution that keep computations tractable.

Axisymmetry could be relaxed by using the Kent distribution, a straightforward generalization of the Fisher distribution, but with elliptical contours (Fisher, Lewis, and Embleton 1987).
But when axisymmetry is broken in astrometry, it is often badly broken.
For example, for GRBs that are observed by two or more well-separated satellites, triangulation based on the burst arrival times at the satellites can supplement the individual-instrument direction uncertainties with tight annular constraints, producing highly non-axisymmetric direction uncertainties.%
\footnote{For more information about such GRB direction constraints from the Interplanetary Gamma-Ray Burst Timing Network (IPN), see the IPN3 web site: \url{http://www.ssl.berkeley.edu/ipn3/}.}
Also, observations of non-electromagnetic ``messengers,'' such as cosmic rays and gravitational waves, can produce complicated likelihood functions because of the complexity of the detector angular response functions; further, the angular resolution for non-electromagnetic detectors is usually much worse than for electromagnetic detectors.

\begin{figure}[t]
  \centering
  \includegraphics[width=.7\textwidth]{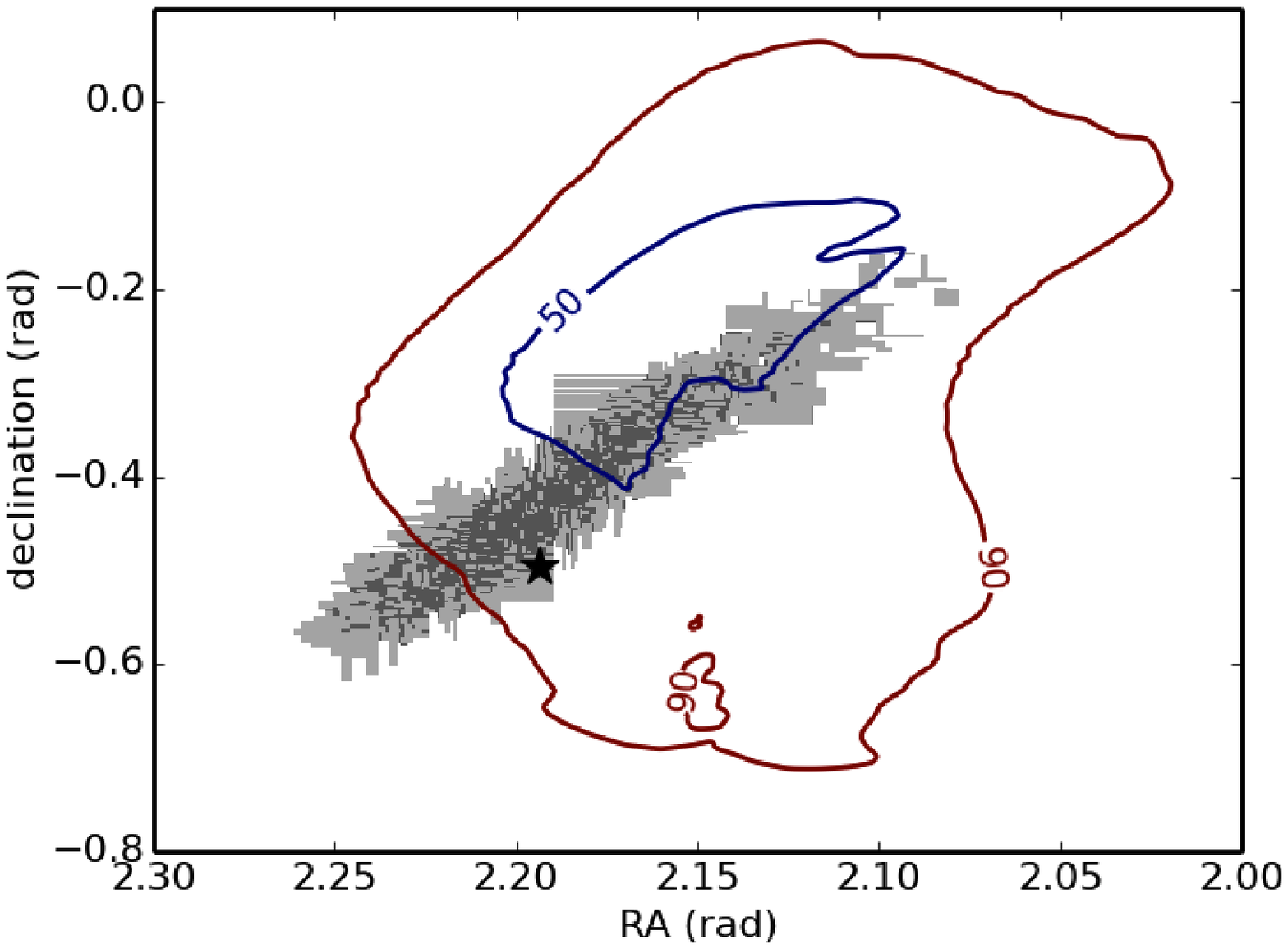}
  \caption{Posterior credible regions for a source direction based on analyses of simulated gravitational wave data (Sidery et al.\ 2014).
  Dark and light gray regions show the 50\% and 90\% credible regions from a computationally expensive algorithm; blue and red contours bound the 50\% and 90\% credible regions using a fast triangulation-based approximate algorithm.
  Star shows the true direction.}
  \label{fig:GWLocn}
\end{figure}

As an illustration, \cfig{fig:GWLocn}\ shows posterior credible regions for a source direction calculated by applying two approximate Bayesian algorithms to simulated gravitational wave data mimicking data from by the two Laser Interferometer Gravitational Observatory (LIGO) detectors operating in the USA, and the Virgo detector operating in Italy (Sidery et al.\ 2014).
The regions are complicated in shape; clearly, a Fisher distribution would poorly represent the uncertainties.
Mixtures of Fisher distributions might provide representations of complex likelihood functions that remain amenable to quick calculation based on closed-form expressions for the marginal likelihood integrals.

The regions in \cfig{fig:GWLocn}\ are not only complicated in shape; they are large, spanning $\approx 0.3$~rad ($\approx 17^\circ$), an area that will contain many stars and galaxies, even restricting attention to nearby objects.
Finding a counterpart for a transient gravitational wave source in so large an area will require \emph{spatio-temporal coincidence assessment}, narrowing candidate counterparts to electromagnetic transient sources coinciding in both direction \emph{and} time with the gravitational wave signal.
GRBs can also have large direction uncertainties, and LLW96 developed a spatio-temporal association model for GRB repetition using the hierarchical framework described here; it requires plausible associations to be comprised of GRBs that have matching directions and that are near to each other in time.
However, computational constraints limited consideration to simplistic models (e.g., with step function temporal behavior), and required crude approximations that were difficult to test.
Spatio-temporal cross-identification needs to be revisited with more modern tools and data.

In the examples described above, we considered situations where the \emph{spatial extent of sources} could be ignored, either because the sources were point-like (stars) or because we expect the source centers to match (galaxies).
But there are settings where the source shape and extent complicates the cross-identification task.
For example, many galaxies have a supermassive black hole at their center.
When the black hole is actively accreting gas and stars from its host galaxy, the nucleus of the galaxy becomes bright, detectable as a point-like active galactic nucleus (AGN, mentioned earlier as a candidate source of UHECRs).
An AGN is often associated with a relativistic jet driving matter and radiation far into intergalactic space, creating lobes and other structures, visible in some wavebands but not others, making multiwavelength active galaxy cross-identification challenging.
To address this, simple source extent models can be included in the hierarchical Bayesian framework, at the cost of extra numerical integrals (Fan, Budav\'ari, and Loredo \ 2014, in preparation).
Further complications arise in situations where the very definition of an object may differ across experiments.
For example, as seen by current X~ray instruments, a galaxy cluster may appear as a single, unresolved source (due both to limited resolution, and to the extended geometry of hot cluster gas emitting X~rays).
Matching an X~ray cluster candidate to an a priori unknown set of galaxies well-resolved in optical or other wavebands is a more difficult task than matching point sources to each other.

Finally, the emerging field of \emph{time-domain synoptic astronomy}---the large-scale study of the temporal behavior of populations of objects---is raising new statistical and computational cross-identification challenges.
The namesake instrument of this field, the Large Synoptic Survey Telescope (LSST, currently under construction), will observe the entire southern sky a thousand times over a decade in several wavebands, generating an avalanche of petabytes of image data, summarized in source catalogs with time series data for hundreds of millions of sources.
To detect the dimmest sources, ideally one would jointly analyze all raw images across all epochs, but this is not feasible.
An alternative strategy is to construct catalogs containing sources of marginal significance at each epoch, and to use those source catalogs as inputs for cross-epoch cross-identification, requiring sources to match, not only in direction, but also in brightness, in order to qualify as a genuine constant-flux object.
In work in progress, we are using the probabilistic framework described in this review to build just such a multi-attribute cross-match algorithm for detecting weak, constant-flux objects with repeated measurements.
How to further generalize the approach, aiming to detect variable and transient sources using synoptic survey catalog data, remains an important open research area.

\section*{ACKNOWLEDGMENTS} 
We are grateful for support from a number of agencies for our work on probabilistic cross-identification across diverse projects, and to many colleagues for insightful contributions to our research.
Budav\'ari gladly acknowledges invaluable discussions over the years on various aspects of cross-identification with Alex Szalay, Andy Connolly, Steve Lubow, Rick White, Bob Mann, Gerard Lemson, Andrew Hopkins and Ray Norris.
L\'aszl\'o Dobos, Gy\"ongyi Kerekes, S\'ebastien Heinis, Matthias Lee, Mar\'{\i}a Jose Marquez and Dongwei Fan helped tremendously with exploring different matching scenarios and creating efficient implementations.
Their efforts were supported by the Gordon and Betty Moore Foundation via grants GBMF 554 and 554.02, as well as by NASA via AISRP grant NNX09AK62G.
NSF provided partial funding for the development of cross-identification tools via NSF grant AST-0122449, Virtual Astronomical Observatory grant VAO\_2010\_06, and as part of the Data Infrastructure Building Blocks (DIBBS) project funded by grant ACI-1261715.
Loredo's early work on cross-identification of GRBs was partially supported
by NASA grant NAG~5-2762; Ira Wasserman helped develop the partition framework, and Shan Luo helped with calculations.
His recent and ongoing work on cross-identification has been supported by interdisciplinary NSF grants AST-0908439 and AST-1312903, in collaboration with David Chernoff, David Ruppert, Kunlaya Soiaporn, and Wasserman.
Loredo gratefully acknowledges very helpful conversations with Merlise Clyde and James Scott that pointed to connections between cross-identification and both mixture models and product partition models.


\nocite{*}

\bibliographystyle{ar-style1}
\newcommand{\apj}{Astrophysical Journal}
\newcommand{\apjl}{Astrophysical Journal Letters}
\newcommand{\aj}{Astronomical Journal}
\newcommand{\mnras}{Monthly Notices of the Royal Astronomical Society}
\newcommand{\jcap}{Journal of Cosmology and Astroparticle Physics}
\newcommand{\aap}{Astronomy \& Astrophysics}
\newcommand{\aaps}{Astronomy \& Astrophysics, Supplement}
\newcommand{\prd}{Physical Review D}
\newcommand{\pasp}{Publications of the Astronomical Society of the Pacific}

% ...

\bibliography{XMATCH}

\begin{thebibliography}{}
\expandafter\ifx\csname natexlab\endcsname\relax\def\natexlab#1{#1}\fi

\bibitem[{{Band} \& {Hartmann}(1998)}]{BH98-GRBHost}
{Band} DL, {Hartmann} DH. 1998.
{A Statistical Treatment of the Gamma-Ray Burst ``No Host Galaxy'' Problem. I.
  Methodology}.
\textit{\apj} 493:555--562

\bibitem[{Berger(2003)}]{B03-FisherLecture}
Berger JO. 2003.
Could {F}isher, {J}effreys and {N}eyman have agreed on testing?
\textit{Statist. Sci.} 18:1--32.
With comments and a rejoinder by the author

\bibitem[{Bernardo \& Gir{\'o}n(1988)}]{BG88-BayesFMM}
Bernardo JM, Gir{\'o}n FJ. 1988.
In \textit{Bayesian statistics, 3 ({V}alencia, 1987)}, Oxford Sci. Publ. New
  York: Oxford Univ. Press,  67--78

\bibitem[{{Budav{\'a}ri} et~al.(2009){Budav{\'a}ri}, {Heinis}, {Szalay},
  {Nieto-Santisteban}, {Gupchup} et~al.}]{B+09-GALEX+SDSS}
{Budav{\'a}ri} T, {Heinis} S, {Szalay} AS, {Nieto-Santisteban} M, {Gupchup} J,
  et~al. 2009.
{GALEX-SDSS Catalogs for Statistical Studies}.
\textit{\apj} 694:1281--1292

\bibitem[{{Budav{\'a}ri} \& {Szalay}(2008)}]{BS08-BayesCrossID}
{Budav{\'a}ri} T, {Szalay} AS. 2008.
{Probabilistic Cross-Identification of Astronomical Sources}.
\textit{\apj} 679:301--309

\bibitem[{{Budav{\'a}ri}, {Szalay} \& {Fekete}(2010)}]{2010PASP..122.1375B}
{Budav{\'a}ri} T, {Szalay} AS, {Fekete} G. 2010.
{Searchable Sky Coverage of Astronomical Observations: Footprints and
  Exposures}.
\textit{\pasp} 122:1375--1388

\bibitem[{Budavári, Dobos \& Szalay(2013)}]{budavari_skyquery:_2013}
Budavári T, Dobos L, Szalay AS. 2013.
{SkyQuery}: Federating astronomy archives.
\textit{Computing in Science \& Engineering} 15:12--20

\bibitem[{Carroll et~al.(2006)Carroll, Ruppert, Stefanski \&
  Crainiceanu}]{CRS06-MsmtErr}
Carroll RJ, Ruppert D, Stefanski LA, Crainiceanu CM. 2006.
Measurement error in nonlinear models, vol. 105 of \textit{Monographs on
  Statistics and Applied Probability}.
Chapman \& Hall/CRC, Boca Raton, FL, 2nd ed.
A modern perspective

\bibitem[{Crowley(1997)}]{C97-PPMNormMeans}
Crowley EM. 1997.
Product partition models for normal means.
\textit{Journal of the American Statistical Association} 92:192--198

\bibitem[{{Dennerl} et~al.(1994){Dennerl}, {Voges}, {Englhauser}, {Gruber},
  {Pfeffermann} et~al.}]{D+94-Orion}
{Dennerl} K, {Voges} W, {Englhauser} J, {Gruber} R, {Pfeffermann} E, et~al.
  1994.
In \textit{Astronomische Gesellschaft Abstract Series}, ed. G~{Klare}, vol.~10
  of \textit{Astronomische Gesellschaft Abstract Series}

\bibitem[{Diaconis \& Mosteller(1989)}]{DM89-Coinc}
Diaconis P, Mosteller F. 1989.
Methods for studying coincidences.
\textit{Journal of the American Statistical Association} 84:853--861

\bibitem[{Fellegi \& Sunter(1969)}]{FS69-RecordLink}
Fellegi IP, Sunter AB. 1969.
A theory for record linkage.
\textit{Journal of the American Statistical Association} 64:1183--1210

\bibitem[{{Fioc}(2014)}]{F14-PosnAssoc}
{Fioc} M. 2014.
{Probabilistic positional association of catalogs of astrophysical sources: the
  Aspects code}.
\textit{\aap} 566:A8

\bibitem[{Fisher, Lewis \& Embleton(1987)}]{FLE87-SphData}
Fisher NI, Lewis T, Embleton BJJ. 1987.
Statistical analysis of spherical data.
Cambridge University Press, Cambridge

\bibitem[{Fisher(1953)}]{F53-FisherDistn}
Fisher R. 1953.
Dispersion on a sphere.
\textit{Proceedings of the Royal Society A: Mathematical, Physical and
  Engineering Sciences} 217:295--305

\bibitem[{Gauss(1809)}]{G1809-Gaussian}
Gauss C. 1809.
Theoria motus corporum celestium: in sectionibus conicis solem ambientium.
I. H. Besser, Hamburg Germany

\bibitem[{{Graziani} \& {Lamb}(1996)}]{GL96-GRBRep}
{Graziani} C, {Lamb} DQ. 1996.
In \textit{High Velocity Neutron Stars}, eds. RE~{Rothschild},
  RE~{Lingenfelter}, vol. 366 of \textit{American Institute of Physics
  Conference Series}

\bibitem[{{Graziani}, {Lamb} \& {Marion}(1999)}]{GLM99-GRB+SNIa}
{Graziani} C, {Lamb} DQ, {Marion} GH. 1999.
{Evidence against an association between gamma-ray bursts and Type I
  supernovae}.
\textit{\aaps} 138:469--470

\bibitem[{Green \& Mardia(2006)}]{green_bayesian_2006}
Green PJ, Mardia KV. 2006.
Bayesian alignment using hierarchical models, with applications in protein
  bioinformatics.
\textit{Biometrika} 93:235--254

\bibitem[{Hartigan(1990)}]{H90-PPMs}
Hartigan J. 1990.
Partition models.
\textit{Communications in Statistics - Theory and Methods} 19:2745--2756

\bibitem[{Kenobi \& Dryden(2012)}]{kenobi_bayesian_2012}
Kenobi K, Dryden IL. 2012.
Bayesian matching of unlabeled point sets using procrustes and configuration
  models.
\textit{Bayesian Analysis} 7:547--566

\bibitem[{{Kerekes} et~al.(2010){Kerekes}, {Budav{\'a}ri}, {Csabai}, {Connolly}
  \& {Szalay}}]{2010ApJ...719...59K}
{Kerekes} G, {Budav{\'a}ri} T, {Csabai} I, {Connolly} AJ, {Szalay} AS. 2010.
{Cross Identification of Stars with Unknown Proper Motions}.
\textit{\apj} 719:59--66

\bibitem[{{Lang} et~al.(2010){Lang}, {Hogg}, {Mierle}, {Blanton} \&
  {Roweis}}]{2010AJ....139.1782L}
{Lang} D, {Hogg} DW, {Mierle} K, {Blanton} M, {Roweis} S. 2010.
{Astrometry.net: Blind Astrometric Calibration of Arbitrary Astronomical
  Images}.
\textit{\aj} 139:1782--1800

\bibitem[{{Lee} \& {Budav{\'a}ri}(2013)}]{2013ASPC..475..235L}
{Lee} MA, {Budav{\'a}ri} T. 2013.
In \textit{Astronomical Data Analysis Software and Systems XXII}, ed.
  DN~{Friedel}, vol. 475 of \textit{Astronomical Society of the Pacific
  Conference Series}

\bibitem[{{LIGO Scientific Collaboration} et~al.(2013){LIGO Scientific
  Collaboration}, {Virgo Collaboration}, {Aasi}, {Abadie}, {Abbott}
  et~al.}]{LIGO+Virgo13-GWLocn}
{LIGO Scientific Collaboration}, {Virgo Collaboration}, {Aasi} J, {Abadie} J,
  {Abbott} BP, et~al. 2013.
{Prospects for Localization of Gravitational Wave Transients by the Advanced
  LIGO and Advanced Virgo Observatories}.
\textit{ArXiv e-prints}

\bibitem[{Loredo(2012)}]{L12-XMatchComment}
Loredo TJ. 2012.
In \textit{Statistical Challenges in Modern Astronomy V}, eds. ED~Feigelson,
  GJ~Babu, Lecture Notes in Statistics. Springer New York,  303--308

\bibitem[{{Luo}, {Loredo} \& {Wasserman}(1996)}]{LLW96-GRBRep}
{Luo} S, {Loredo} T, {Wasserman} I. 1996.
In \textit{American Institute of Physics Conference Series}, eds.
  C~{Kouveliotou}, MF~{Briggs}, GJ~{Fishman}, vol. 384 of \textit{American
  Institute of Physics Conference Series}

\bibitem[{Mardia(1972)}]{M72-StatDrxnData}
Mardia KV. 1972.
Statistics of directional data.
Academic Press, London-New York.
Probability and Mathematical Statistics, No. 13

\bibitem[{{Marquez}, {Budav{\'a}ri} \& {Sarro}(2014)}]{2014A&A...563A..14M}
{Marquez} MJ, {Budav{\'a}ri} T, {Sarro} LM. 2014.
{Improving cross-identification of galaxies using their photometry}.
\textit{\aap} 563:A14

\bibitem[{Sadinle(2014)}]{S14-DupDtxn}
Sadinle M. 2014.
Detecting duplicates in a homicide registry using a bayesian partitioning
  approach.
\textit{Ann. Appl. Stat.} :to appear

\bibitem[{Sadinle \& Fienberg(2013)}]{SF13-RecLink}
Sadinle M, Fienberg SE. 2013.
A generalized {F}ellegi-{S}unter framework for multiple record linkage with
  application to homicide record systems.
\textit{J. Amer. Statist. Assoc.} 108:385--397

\bibitem[{Scott \& Berger(2006)}]{SB06-BayesMultTest}
Scott JG, Berger JO. 2006.
An exploration of aspects of {B}ayesian multiple testing.
\textit{J. Statist. Plann. Inference} 136:2144--2162

\bibitem[{{Sidery} et~al.(2014){Sidery}, {Aylott}, {Christensen}, {Farr},
  {Farr} et~al.}]{S+14-GWLocn}
{Sidery} T, {Aylott} B, {Christensen} N, {Farr} B, {Farr} W, et~al. 2014.
{Reconstructing the sky location of gravitational-wave detected compact binary
  systems: Methodology for testing and comparison}.
\textit{\prd} 89:084060

\bibitem[{Soiaporn et~al.(2013)Soiaporn, Chernoff, Loredo, Ruppert \&
  Wasserman}]{S+13-UHECR}
Soiaporn K, Chernoff D, Loredo T, Ruppert D, Wasserman I. 2013.
Multilevel {B}ayesian framework for modeling the production, propagation and
  detection of ultra-high energy cosmic rays.
\textit{Ann. Appl. Stat.} 7:1249--1285

\bibitem[{Steorts, Hall \& Fienberg(2014)}]{SHF14-BayesRecLink}
Steorts R, Hall R, Fienberg S. 2014.
In \textit{Proceedings of the Seventeenth International Conference on
  Artificial Intelligence and Statistics}

\bibitem[{{Sutherland} \& {Saunders}(1992)}]{1992MNRAS.259..413S}
{Sutherland} W, {Saunders} W. 1992.
{On the likelihood ratio for source identification}.
\textit{\mnras} 259:413--420

\bibitem[{Tancredi \& Liseo(2011)}]{GL11-HBRecLink}
Tancredi A, Liseo B. 2011.
A hierarchical {B}ayesian approach to record linkage and population size
  problems.
\textit{Ann. Appl. Stat.} 5:1553--1585

\bibitem[{{Watson}, {Mortlock} \& {Jaffe}(2011)}]{2011MNRAS.418..206W}
{Watson} LJ, {Mortlock} DJ, {Jaffe} AH. 2011.
{A Bayesian analysis of the 27 highest energy cosmic rays detected by the
  Pierre Auger Observatory}.
\textit{\mnras} 418:206--213

\end{thebibliography}

\end{document}